\documentclass[iop,revtex4]{emulateapj}

\let\pwiflocal=\iffalse \let\pwifjournal=\iffalse

\usepackage{amsmath,amssymb}
\usepackage[breaklinks,colorlinks,urlcolor=blue,citecolor=blue,linkcolor=blue]{hyperref}
\usepackage{epsfig}    
\usepackage{graphicx}    
\usepackage{lineno}
\usepackage{natbib}
\usepackage{bigints}
\usepackage[outdir=./]{epstopdf}

\usepackage[T1]{fontenc}
\pwifjournal\else
  \usepackage{microtype}
\fi

\pwifjournal\else
  \makeatletter
  \renewcommand\plotone[1]{%
    \centering \leavevmode \setlength{\plot@width}{0.95\linewidth}
    \includegraphics[width={\eps@scaling\plot@width}]{#1}%
  }%
  \makeatother
\fi

\makeatletter

\newcommand\@simpfx{http://simbad.u-strasbg.fr/simbad/sim-id?Ident=}

\newcommand\MakeObj[4][\@empty]{
  \pwifjournal%
    \expandafter\newcommand\csname pkgwobj@c@#2\endcsname[1]{\protect\object[#4]{##1}}%
  \else%
    \expandafter\newcommand\csname pkgwobj@c@#2\endcsname[1]{\href{\@simpfx #3}{##1}}%
  \fi%
  \expandafter\newcommand\csname pkgwobj@f#2\endcsname{#4}%
  \ifx\@empty#1%
    \expandafter\newcommand\csname pkgwobj@s#2\endcsname{#4}%
  \else%
    \expandafter\newcommand\csname pkgwobj@s#2\endcsname{#1}%
  \fi}%

\newcommand\MakeTrunc[2]{
  \expandafter\newcommand\csname pkgwobj@t#1\endcsname{#2}}%

\newcommand{\obj}[1]{%
  \expandafter\ifx\csname pkgwobj@c@#1\endcsname\relax%
    \textbf{[unknown object!]}%
  \else%
    \csname pkgwobj@c@#1\endcsname{\csname pkgwobj@s#1\endcsname}%
  \fi}
\newcommand{\objf}[1]{%
  \expandafter\ifx\csname pkgwobj@c@#1\endcsname\relax%
    \textbf{[unknown object!]}%
  \else%
    \csname pkgwobj@c@#1\endcsname{\csname pkgwobj@f#1\endcsname}%
  \fi}
\newcommand{\objt}[1]{%
  \expandafter\ifx\csname pkgwobj@c@#1\endcsname\relax%
    \textbf{[unknown object!]}%
  \else%
    \csname pkgwobj@c@#1\endcsname{\csname pkgwobj@t#1\endcsname}%
  \fi}

\makeatother

\pwifjournal\else
  \usepackage{etoolbox}
  \makeatletter
  \patchcmd{\NAT@citex}
    {\@citea\NAT@hyper@{%
       \NAT@nmfmt{\NAT@nm}%
       \hyper@natlinkbreak{\NAT@aysep\NAT@spacechar}{\@citeb\@extra@b@citeb}%
       \NAT@date}}
    {\@citea\NAT@nmfmt{\NAT@nm}%
     \NAT@aysep\NAT@spacechar\NAT@hyper@{\NAT@date}}{}{}
  \patchcmd{\NAT@citex}
    {\@citea\NAT@hyper@{%
       \NAT@nmfmt{\NAT@nm}%
       \hyper@natlinkbreak{\NAT@spacechar\NAT@@open\if*#1*\else#1\NAT@spacechar\fi}%
         {\@citeb\@extra@b@citeb}%
       \NAT@date}}
    {\@citea\NAT@nmfmt{\NAT@nm}%
     \NAT@spacechar\NAT@@open\if*#1*\else#1\NAT@spacechar\fi\NAT@hyper@{\NAT@date}}
    {}{}
  \makeatother
\fi

\MakeObj{n33370}{NLTT\%2033370}{NLTT~33370\,AB}
\MakeTrunc{n33370}{NLTT~33370}
\MakeObj{tvlm}{TVLM\%20513-46546}{TVLM~513--46546}

\newcommand{\kepler}{{\it Kepler}}

\newcommand{\um}{$\mu$m}

\newcommand\kms{km~s$^{-1}$}

\newcommand\ms{M$_\odot$}

\newcommand\teff{\ensuremath{T_\text{eff}}}

\providecommand{\adsurl}[1]{\href{#1}{ADS}}
\newcommand{\name}{K2-25}
\newcommand{\pname}{K2-25b}
\newcommand{\prad}{$ 3.43^{+0.95}_{-0.31}$\,$R_\earth$}
\newcommand{\mps}{m\,s$^{-1}$}
\providecommand{\icarus}{Icarus}
\def\vsini{$v\sin{i_*}$}

\slugcomment{Received November 23, 2015; Accepted December 17, 2015}

\shorttitle{A Transiting Planet in the Hyades}

\shortauthors{Mann et al.}

\begin{document}
 
\title{Zodiacal Exoplanets in Time (ZEIT). I.\\ A Neptune-sized planet orbiting an M4.5 dwarf in the Hyades Star Cluster}

\author{Andrew W. Mann,\altaffilmark{$\star$,1,2} Eric Gaidos,\altaffilmark{3,4} Gregory N. Mace,\altaffilmark{2} Marshall C. Johnson,\altaffilmark{2} Brendan P. Bowler,\altaffilmark{2,5,6} Daryll LaCourse,\altaffilmark{7} Thomas L. Jacobs\altaffilmark{7}, Andrew Vanderburg,\altaffilmark{8,9} Adam L. Kraus,\altaffilmark{2} Kyle F. Kaplan,\altaffilmark{2} Daniel T. Jaffe\altaffilmark{2}}

\altaffiltext{$\star$}{amann@astro.as.utexas.edu}
\altaffiltext{1}{Hubble Fellow}
\altaffiltext{2}{Department of Astronomy, The University of Texas at Austin, Austin, TX 78712, USA}
\altaffiltext{3}{Visiting Astronomer at the Infrared Telescope Facility, which is operated by the University of Hawaii under contract NNH14CK55B with the National Aeronautics and Space Administration}
\altaffiltext{4}{Department of Geology \& Geophysics, University of Hawaii at Manoa, Honolulu, HI 96822, USA}
\altaffiltext{5}{California Institute of Technology, 1200 E. California Blvd., Pasadena, CA 91125, USA.}
\altaffiltext{6}{McDonald Fellow}
\altaffiltext{7}{Amateur Astronomer}
\altaffiltext{8}{Harvard-Smithsonian Center for Astrophysics, Cambridge, MA 02138, USA}
\altaffiltext{9}{NSF Graduate Research Fellow}

\begin{abstract}
Studying the properties of young planetary systems can shed light on how the dynamics and structure of planets evolve during their most formative years. Recent {\it K2} observations of nearby young clusters (10-800 Myr) have facilitated the discovery of such planetary systems. Here we report the discovery of a Neptune-sized planet transiting an M4.5 dwarf (\name) in the Hyades cluster (650-800\,Myr). The light curve shows a strong periodic signal at 1.88\,days, which we attribute to spot coverage and rotation. We confirm that the planet host is a member of the Hyades by measuring the radial velocity of the system with the high-resolution near-infrared spectrograph IGRINS. This enables us to calculate a distance based on \name's kinematics and membership to the Hyades, which in turn provides a stellar radius and mass to $\simeq5-10\%$, better than what is currently possible for most \kepler{} M dwarfs (12-20\%). We use the derived stellar density as a prior on fitting the {\it K2} transit photometry, which provides weak constraints on eccentricity. Utilizing a combination of adaptive optics imaging and high-resolution spectra, we rule out the possibility that the signal is due to a bound or background eclipsing binary, confirming the transits' planetary origin. \pname\ has a radius (\prad) much larger than older \kepler{} planets with similar orbital periods (3.485\,days) and host-star masses (0.29\ms). This suggests that close-in planets lose some of their atmospheres past the first few hundred million years. Additional transiting planets around the Hyades, Pleiades, and Praesepe clusters from {\it K2} will help confirm whether this planet is atypical or representative of other close-in planets of similar age. 

\end{abstract}


\maketitle

\section{Introduction}\label{sec:intro}

Planets and their host stars evolve with time, and the first few hundred million years are thought to be the most formative. Final assembly of rocky terrestrial planets is predicted to occur in 10-100~Myr \citep{Morbidelli2012}. Regular accretion of residual planetesimals would continue to influence physical and chemical conditions on these planets \citep{Hashimoto2007,Abramov2009}. More rapid rotation and magnetic activity drive elevated X-ray and ultraviolet emission and coronal mass ejections from the host star, potentially eroding the primordial atmospheres of close-in planets on this timescale \citep{Lammer2014}. 

M dwarfs play a disproportionately large role in the discovery of Earth-size planets, particularly those planets with theoretical equilibrium temperatures permissive of liquid water \citep{Muirhead2012,Dressing2013,Gaidos2013a,Mann2013,Dressing2015}. This is because M dwarfs have smaller radii than solar-type stars, permitting the detection of smaller planets, and much lower luminosities, such that close-in and detectable planets will also be cooler. The dynamical and structural evolution of these systems may be qualitatively different than that of Sun-like stars; M dwarf stars take much longer ($\sim 10^8$~yr) to drop onto the main sequence and remain active much longer than their solar-type counterparts \citep{Ansdell2015, West2015}. These characteristics could have consequences for the climatic states and atmospheric evolution of M dwarf planets \citep{Luger2015}. 

While thousands of exoplanets have been discovered, most by the NASA \kepler\ transiting planet survey mission \citep{Borucki2010}, the vast majority of these orbit old ($\gg1$\,Gyr) stars. However, the repurposed \kepler{} spacecraft \citep[{\it K2},][]{Howell2014} has observed 10-800 Myr-old clusters (i.e., Upper Scorpius, Pleiades, Hyades, and Praesepe). The TESS \citep{Ricker2014} and {\it PLATO} \citep{Rauer2014} missions will also observe many stars in both young clusters and nearby young moving groups. These surveys will populate the temporal dimension of exoplanet parameter space, allowing us to statistically deduce how their orbits, masses, and atmospheres change with time.

Until catalogs from the {\it Gaia} mission become available, stars in well-studied clusters are usually easier to characterize than their counterparts in the field. Precise abundances can be determined from the Sun-like members and can be applied to late-type stars where abundance determinations are more complicated. The distances to most of the nearest young clusters are well established \citep{van-Leeuwen2009, 2014Sci...345.1029M}. While one cannot use the exact cluster distance for individual members of large clusters like the Hyades that have members $>20$\,pc from the cluster center, it is still possible to derive an accurate (5-10\%) "kinematic distance," i.e., the distance that yields Galactic kinematics ($UVW$) consistent with the cluster or moving group \citep[e.g.,][]{Roser2011,Malo2013}. For M dwarfs the combination of distance, flux, and temperature can yield radius estimates accurate to 5-10\% \citep{Delfosse2000,Bayless2006,Mann2015b}. This is significantly better than the $13-15\%$ errors for most \kepler{} M dwarfs based on spectroscopy \citep[e.g.,][]{Muirhead2014,Newton2015A} and is less subject to systematic biases that have plagued stellar characterization of \kepler{} targets from photometry alone \citep[e.g.,][]{Mann:2012,Gaidos:2013qy,Gaidos2013b}.  Perhaps more significant, the common {\it age} of all members can be established by applying different methods (e.g., asteroseismology, isochrone fitting, lithium depletion boundary) to the full set of stars.

A handful of planets have already been discovered around stars in young open clusters, i.e., by the Doppler radial velocity (RV) method in the Hyades (650-800\,Myr) and Praesepe (650-800\,Myr) clusters  \citep{2007ApJ...661..527S,2012ApJ...756L..33Q,Quinn2014} and by \kepler\ in the $\sim$1~Gyr old NGC~6811 cluster \citep{Meibom2013}. Owing to the sensitivity of the Doppler method, the former sample is limited to Jupiter-mass planets, which represent only a tiny fraction of the total planet population. The latter is limited to two Neptune-sized planets in a single, distant ($\approx 1100$~kpc, and therefore difficult to study) cluster. {\it K2} is observing much closer clusters with a range of ages and has already proved  precise enough to find close-in ($P\lesssim20$~days), Neptune-size and smaller planets \citep[e.g.,][]{2015ApJ...804...10C,2015ApJ...806..215F,Petigura2015a, 2015ApJ...800...59V}

Here we present the {\it K2} discovery and our validation and characterization of a Neptune-sized planet transiting a mid-M-type dwarf (EPIC~210490365, 2MASS J04130560+1514520, \name) in the Hyades cluster.  This object was identified by visual inspection of the host star's {\it K2} light curve shortly after public release.  In Section~\ref{sec:obs} we describe our spectroscopic and imaging follow-up, as well as literature photometry of the host star and extraction of the {\it K2} light curve. We use these data and others from the literature in Section~\ref{sec:hyades} to show that this is a true member of the Hyades cluster. Using the membership status of \name, we derive a kinematic distance to \name, which in turn we utilize in Section~\ref{sec:params} to derive accurate stellar parameters of the star. Our fit to the transit light curve is described in Section~\ref{sec:transit}. In Section~\ref{sec:fpp} we combine our stellar and planetary parameters with our adaptive optics (AO) and high-resolution observations to confirm the planetary nature of this transit. In Section~\ref{sec:discussion} we conclude with a brief summary and discussion of the tentative implications for this system, the key differences between this planet and those found by \kepler{} around old M dwarfs, and the need for additional follow-up.

\section{Archival and Follow-up Observations}\label{sec:obs} 

\subsection{Archival Photometry/Imaging}\label{sec:phot}
We compiled optical $BV$ photometry from the eighth data release of the AAVSO All-Sky Photometric Survey \citep[APASS;][]{Henden:2012fk}, NIR $JHK_S$ photometry from The Two Micron All Sky Survey \citep[2MASS;][]{Skrutskie2006}, $griz$ photometry from the Sloan Digital Sky Survey \citep[SDSS;][]{Ahn:2012kx}, and $W1W2W3$ infrared photometry from the Wide-field Infrared Survey Explorer \citep[WISE;][]{Wright:2010fk}. We retrieved proper motions for \name\ from \citet{Roser2011}, which combined PPMXL \citep{2010AJ....139.2440R} and UCAC3 \citep{Zacharias:2010lr} proper motions. These basic data on \name\ are given in Table~\ref{tab:params}.

The SDSS images are sufficiently deep to detect faint background stars ($r'<21$) that might fall within the {\it K2} aperture (provided they are $\gg1\arcsec$\ from the star) and contaminate the light curve or generate a false positive (if they are eclipsing binaries). The DSS image complements this, by offering a view directly behind the star when the target was $\sim7\arcsec$\ away owing to its large proper motion between the epoch of the Palomar Observatory Sky Survey (POSS, 1953) and {\it K2} (2015) observations. We utilized these images as part of our false-positive analysis described in Section~\ref{sec:fpp}.

\subsection{{\it K2} light curve}\label{sec:lc}
\name\ was observed by the repurposed \kepler{} satellite \citep[{\it K2},][]{Howell2014} for $\simeq$71 days from 2015 February 8 to April 20. Telescope pointing for {\it K2} is unstable due to the loss of two reaction wheels, so the telescope drifts slowly. When the roll angle deviates too far from the desired position the thrusters fire to correct the pointing. As the point-spread function of the star shifts during the roll and subsequent thrust the total measured flux from the star changes due to pixel-to-pixel sensitivity variations. The resulting systematic noise is on timescales matching the thrust and roll ($\sim$6 hr). Fortunately, these deviations can be corrected for or at least mitigated \citep[][]{Vanderburg2014,2015A&A...582A..33A}. We retrieved the light curve of \name\ provided by \citet{Vanderburg2014}, who generate light curves from {\it K2} pixel data accounting for the nonuniform pixel response function by fitting out correlations between the flux levels and spacecraft pointing. We also downloaded the light curve provided by the \kepler{} and {\it K2} Science Center to test our sensitivity to the corrections applied by \citet{Vanderburg2014}.

\subsection{Optical Spectrum}
We obtained an optical spectrum of \name\ with the SuperNova Integral Field Spectrograph \citep[SNIFS;][]{Aldering2002,Lantz2004} on the University of Hawaii 2.2m telescope on Mauna Kea. SNIFS provides simultaneous coverage from 3200 to 9700\,\AA\ by splitting the beam with a dichroic mirror onto blue (3200--5200\,\AA) and red (5100--9700\,\AA) spectrograph channels. Although the spectral resolution of SNIFS is only $R\simeq1000$, the instrument provides excellent spectrophotometric precision \citep{Mann:2011qy, Buton:2013}. Three optical spectra were taken of the target in succession (each integrated for 900\,s) under cloudy conditions. After reduction the three spectra were stacked to yield a signal-to-noise ratio (S/N) of $\simeq$200 (per pixel) in the $r$ band.

Reduction of SNIFS data was split into two parts; the first part was done by the SuperNova Factory (SNF) pipeline, which performs basic reduction (e.g., bias and flat correction) and extraction of the 1-dimensional spectrum.  The second section, carried out by our pipeline. applies flux and telluric correction and places the star in its rest frame. Extensive details of the SNF pipeline can be found in \citet{Bacon:2001} and \citet{Aldering:2006}; and details of our pipeline can be found in \citet{Gaidos2014} and \citet{Mann2015b}. 

\subsection{Near-Infrared Spectrum}

A near-infrared (NIR) spectrum was taken with the updated SpeX spectrograph \citep{Rayner2003} mounted on the NASA Infrared Telescope Facility (IRTF) on Mauna Kea. SpeX observations were taken in the short cross-dispersed (SXD) mode using the 0.3$\times15\arcsec$ slit, yielding simultaneous coverage from 0.8 to 2.4\um\ at a resolution of $R\simeq2000$. The target was placed at two positions along the slit (A and B) and observed in an ABBA pattern in order to subsequently subtract the sky background. We took 6 exposures following this pattern for a total integration time of 717s, which, when stacked, provided a S/N per pixel of 120 in the $H$ and $K$ bands.

SpeX spectra were extracted using the SpeXTool package \citep{Cushing2004}, which performed flat-field correction, wavelength calibration, sky subtraction, and extraction of the one-dimensional spectrum. Multiple exposures were combined using the IDL routine \textit{xcombspec}. To correct for telluric lines, we observed the A-type star HD\,31295 immediately after the target observations and within 0.1\,airmasses. A telluric correction spectrum was constructed from the A0V star and applied using the \textit{xtellcor} package \citep{Vacca2003}. Separate orders were stacked using the \textit{xcombspec} tool.

\begin{figure*}
	\centering
	\includegraphics[width=0.95\textwidth]{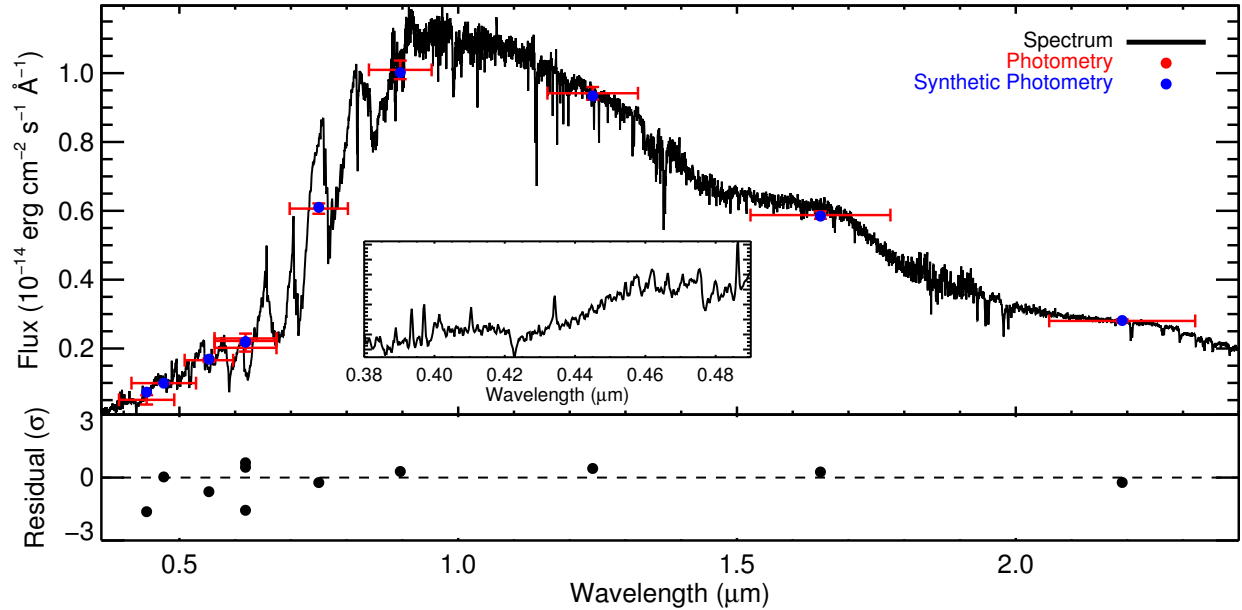} 
	\caption{Combined and absolutely flux-calibrated optical and NIR spectrum of \name. The spectrum is shown in black. Photometry is shown in red, with the horizontal "error bars" indicating the width of the filter, and vertical errors representing combined measurement and zero point errors. Blue points indicate the corresponding synthetic fluxes from convolving the spectrum with the appropriate filter profile and multiplying by the zero point. Residuals are plotted in the bottom panel in units of standard deviations. The inset panel shows a zoom-in of the blue part of the spectrum including the major Balmer and calcium H \& K emission lines.}
	\label{fig:spec}
\end{figure*}

Following the method outlined in \citet{Mann2015b} we combined and absolutely flux-calibrated the optical and NIR spectra using published photometry (Section~\ref{sec:phot}) with the filter profiles and zero points provided in \citet{Fukugita:1996}\footnote{See \href{http://classic.sdss.org/dr7/algorithms/fluxcal.html}{http://classic.sdss.org/dr7/algorithms/fluxcal.html}} and \citet{Mann2015a}. We show the combined spectrum in Figure~\ref{fig:spec}.

\subsection{High Resolution NIR Spectra}\label{sec:igrins}%

We observed \name\ at 10 epochs spread over 36 days with the Immersion Grating Infrared Spectrometer \citep[IGRINS;][]{Park2014} on the 2.7m Harlan J. Smith telescope at McDonald Observatory. IGRINS provides simultaneous $H$- and $K$-band (1.48-2.48\um) coverage with a resolving power of $R\simeq$45,000. Similar to the SpeX observations, the target was placed at two positions along the slit and observed in an ABBA pattern. At each epoch we took four exposures (one ABBA cycle), each 240-400s (depending on conditions). For each epoch we also observed an A0V star immediately before or after the observations of \name\ to aid with telluric correction.

IGRINS spectra were reduced using version 2.1 of the publicly available IGRINS pipeline package\footnote{https://github.com/igrins/plp} \citep{jae_joon_lee_2015_18579}. The IGRINS pipeline performed flat, bias, and dark corrections, as well as extraction of the one-dimensional spectrum of both the A0V standard and target. An initial wavelength solution was derived using the ThAr lines, followed by a full wavelength solution derived from the sky lines. The resulting spectrum at each epoch has a S/N of $30-60$ per pixel in the center of the $H$ and $K$ bands. We used the A0V spectra to correct for telluric lines following the method outlined in \citet{Vacca2003}. Spectra with uncorrected telluric lines were used for measuring radial velocities to improve the wavelength solution (see Section~\ref{sec:hyades}). A single high-S/N spectrum was constructed by stacking the ten exposures after shifting them to the same radial velocity. The final stacked spectrum has a S/N of $\sim120$ in the $H$-band, which we used to search for faint lines from an undetected companion (Section~\ref{sec:fpp}) and calculate \vsini\ (Section~\ref{sec:params}).

\subsection{Adaptive Optics Imaging} \label{sec:ao}

\begin{figure*}
	\centering
	\includegraphics[trim={3.8cm 7.5cm 1.6cm 2.0cm},,clip,width=0.95\textwidth]{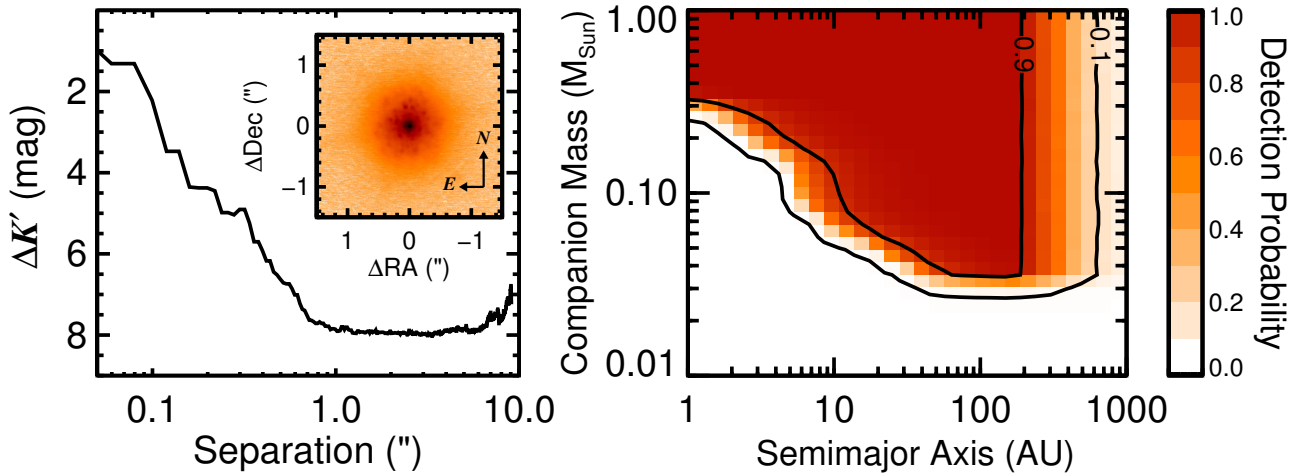} 
	\caption{7-$\sigma$ contrast curve (left) and sensitivity map (right) for \name\ constructed from our AO imaging. The registered and stacked AO image is shown as an inset on the left panel. The sensitivity map is a measure of the probability of detecting an object of a given mass and separation based on an age of 650\,Myr and the distance to \name\ (Section~\ref{sec:params}). See \citet{Bowler2015a} for more details. }
	\label{fig:AO}
\end{figure*}

We obtained natural guide star \citep[NGS;][]{2000PASP..112..315W} AO imaging of \name\ with the facility imager, NIRC2, on Keck II atop Mauna Kea. Observations were taken with the $K'$ filter and the narrow camera. In this mode the pixel scale is 9.952\,mas\,pix$^{-1}$ \citep{Yelda2010} and the field of view (FOV) for the $1024\times1024$ pixel array is $10.2\arcsec \times 10.2\arcsec$. We took seven images, each with five co-adds, and each co-add integrating for 2\,s. Basic reduction (dark, flat field, and bad pixel correction) was applied to each of the images, which were then registered and stacked to produce a single, deep image. No sources are visible in the fFOV other than \name\ down to the resolution limit of the images ($\simeq0.07\arcsec$). From the reduced and stacked images we constructed a contrast curve using the noise maps and the flux from the primary star following \citet{Bowler2015a}, which we show in Figure~\ref{fig:AO}.

Using the age of the 650\,Myr for Hyades, with the $K_S$ magnitude and distance of \name (Section~\ref{sec:params}), and the models of \citet{2015A&A...577A..42B}, we converted our AO image into a sensitivity map following the procedure from \citet{Bowler2015a}, which we show in Figure~\ref{fig:AO}. The result does not significantly change for an age of 800\,Myr. This provides a probability of detecting a star of a given separation and mass in the AO images and accounts for the chance of detecting a companion at a random place in its orbit.

\section{Hyades Membership}\label{sec:hyades}

\name\ was previously identified as a member of the Hyades in \citet{Roser2011} based on its proper motion and photometry. Using the subset of their candidates with parallaxes and a comparison of the density of the Hyades and field stars, \citet{Roser2011} estimate that field star contamination within 9\,pc of the cluster core is negligibly small, and they find \name\ to be only 3.5\,pc from the core. Similarly, \citet{2014ApJ...795..161D} calculate a 99\% chance that \name\ is a member of the Hyades. Furthermore, our high-resolution NIR spectra (Section~\ref{sec:igrins}), combined with the proper motion (Section~\ref{sec:phot}), enable a precise determination of this star's kinematics and hence unambiguously confirm \name's membership in the Hyades.

We calculated a barycentric RV of \name\ for each of the 10 IGRINS epochs following the method from Mace et al. (2016, in preparation), which takes advantage of the large spectral grasp and stability of IGRINS. The procedure was to; (1) split each of 42 orders into eight suborders and remove the two on each end where the S/N is lowest owing to the drop in the blaze function, (2) cross-correlate the telluric spectrum to find offsets in the wavelength solution due to temperature changes and instrument flexures, (3) invert and cross-correlate the remaining telluric-corrected suborders against a template (or series of templates) to determine the offset in pixels, (4) convert the pixel offsets into RVs using the instrument dispersion solution, (5) adopt the median of the RV measurement after removing $>4\sigma$ outliers, and (6) repeat steps 2-4 for every available template with a known RV, allowing the RV of each template to adjust according to their literature uncertainties. In total we compared each observation with 153 M2--M6 templates with known RVs. For each epoch the assigned RV and error is the mean and standard deviation of the mean for these 153 measurements. We use the mean of the RVs from the 10 epochs as the system RV. Nominally the statistical error on the final barycentric RV is $<50$\,\mps, but it is limited by the 153\,\mps\ error in the zero point velocity derived from the templates. Our final barycentric velocity is $38.64\pm0.15$\kms\ and is reported in Table~\ref{tab:params}. 

We estimated a photometric distance to the star by comparing the spectral energy distribution of \name\ (Section~\ref{sec:phot}) with that of template M dwarfs with known distances from \citet{Mann2015b}. For each star in \citet{Mann2015b} we calculated the reduced $\chi^2$ ($\chi^2_\nu$) between the template and \name's $BV$\,$griz$\,$JHK_s$ photometry with the distance of \name\ as a free parameter. We calculated the best-fit distance for \name\ from the median and standard deviation of the 10 stars with $\chi^2_\nu<3$. This yielded a distance of $44\pm7$\,pc, assuming that the star is not an unresolved binary. We used the coordinates, proper motion, RV, and photometric distance to calculate Galactic position $XYZ$ and motion $UVW$ with corresponding errors, which we report in Table~\ref{tab:params}.

We calculated the probability that \name\ is member of the Hyades following the Bayesian framework from \citet{2011MNRAS.416.3108R} and \citet{Malo2013}. For simplicity we only considered the possibility that this star is part of the Hyades or the field. We adopted $XYZ$ and $UVW$ for the Hyades from \citet{van-Leeuwen2009} and the $XYZ$ and $UVW$ values for field stars from \citet{Malo2013}. For the prior \citet{Malo2013} give equal weights to membership in each of the memberships considered; however, this is overly optimistic for our case as there are far more field stars than members of the Hyades. Instead we followed \citet{2011MNRAS.416.3108R} and selected a prior equal to the ratio of the number of stars in the Hyades to the number of field stars in the same region of the sky. We identified all stars within 10$^\circ$ of the target in APASS that land within $5\sigma$ of the color-magnitude diagram of stars in the Hyades drawn from the \citet{Roser2011} and \citet{2013A&A...559A..43G} catalogs. We conservatively assume that all of these stars not in \citet{Roser2011} and \citet{2013A&A...559A..43G} are field stars. 

Plugging the $XYZ$ and $UVW$ values and errors for our target and the Hyades, along with our prior estimated above, into Equation~8 of \citet{Malo2013} gives a 99.99\% chance that \name\ is a member of the Hyades as opposed to a field star. This is relatively insensitive to our choice of prior; we would have to decrease the prior by more than an order of magnitude to drop the membership probability below 99.7\% (3$\sigma$). Our analysis does not consider the chance that \name\ is a member of another moving group, stream, or cluster (only the Hyades and the field). However, the $XYZUVW$ values of \name\ are not consistent with any other known moving group or cluster, so we consider this possibility to have negligible probability. 

\begin{figure*}
	\centering
	\includegraphics[width=0.95\textwidth]{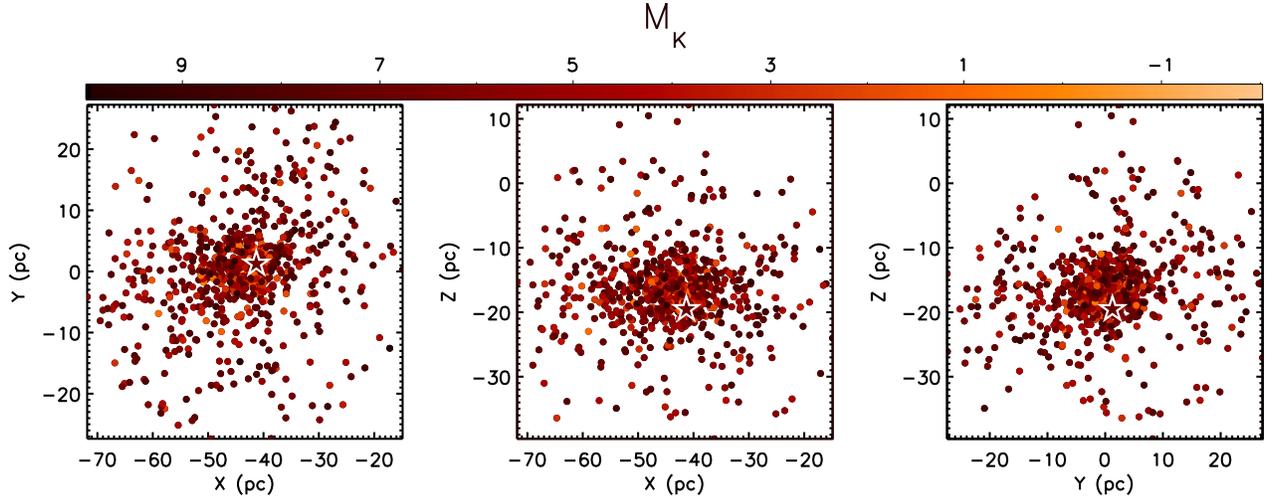} 
	\caption{Galactic position ($XYZ$) of Hyades members (circles) and \name\ (five-point star). All points are colored according to their $M_{K_S}$ magnitude. The member list, as well as distances and coordinates used in computing $XYZ$, was taken from \citet{Roser2011} and \citet{2013A&A...559A..43G}. The background contamination rate (false members) is expected to be low ($<$7.5\%) near the core ($<18$\,pc), but significantly higher ($>30\%$) for more distant stars.}
	\label{fig:xyz}
\end{figure*}

\name\ lands only $\simeq3.4$\,pc from the core of the Hyades and therefore is probably still gravitationally bound to the cluster. We show the Galactic position ($XYZ$) for members of the Hyades taken from \citet{Roser2011} and \citet{2013A&A...559A..43G} compared to that of \name\ in Figure~\ref{fig:xyz}. 

\section{Stellar Parameters}\label{sec:params} 

{\it Spectral type}: We calculated TiO, CaH, and VO molecular indices following the definitions from \citet{Reid:1995lr} and \citet{Lepine:2013}. We then derived a decimal spectral type using the empirical relations between the strength of these molecular indices and the spectral type from \citet{Lepine:2013}. This gives a final spectral type of M4.5 with an internal error of $\le$0.3 subtypes. 

{\it H$\alpha$}: The optical spectrum shows noticeable Balmer series and Calcium H \& K emission (Figure~\ref{fig:spec}), as expected for a mid-M dwarf in the Hyades \citep{2008AJ....135..785W}. We calculated an H$\alpha$ equivalent width of $-3.1$\,\AA\ (negative to denote emission), following the continuum and feature definitions from \citet{Lepine:2013}. The resulting value is consistent with other H$\alpha$ measurements of stars in Hyades and Praesepe clusters \citep{2014ApJ...795..161D}.

{\it Metallicity}: Our SpeX spectrum enables us to derive the host star's metallicity from atomic indices in the $H$ and $K$ bands \citep[e.g., ][]{Terrien:2012lr,Rojas-Ayala:2012uq,Mann2013a,Newton:2014}. However, the commonly used Na lines at 2.2\um\ could be affected by stellar activity and lower surface gravity \citep{2013AJ....146...51D}, although this effect should be smaller for the Ca and K lines in the $H$-band \citep{Terrien:2012lr}. The $H$-band indices from \citet{Terrien:2012lr} and \citet{Mann2013a} give consistent values ([Fe/H] = 0.15$\pm$0.10, 0.17$\pm$0.08) and both are in agreement with literature values for the cluster \citep[0.12--0.18,][]{2003AJ....125.3185P,2015ApJ...807...24B,Dutra-Ferreira2016}. For our analysis we adopted [Fe/H]=0.15$\pm$0.03, which captures both our measurements and those from the literature within 1$\sigma$ and is far more precise than can be done on an individual M dwarf. 

{\it Distance}: Because \name\ is a member of the Hyades, we can derive a kinematic distance more precise than the photometric distance we used in Section~\ref{sec:hyades} \citep[e.g.,][]{Kraus2014}. To this end, we recalculated $UVW$ but allowing distance to float between 1 and 100\,pc. We then found the distance that gives $UVW$ values consistent with the established value of the cluster \citep{van-Leeuwen2009}. We allowed for a variation of 1.2\,km/s in the cluster value due to dispersion from internal kinematics \citep{2014A&A...564A..49P}. Accounting for this, as well as errors in the proper motion and RV, gave a distance of 45.7$\pm3.3$\,pc.  

{\it Luminosity}: We first calculated the bolometric flux by integrating over our combined and absolutely flux-calibrated optical and NIR spectrum (Section~\ref{sec:obs}). This gives a bolometric flux of $1.301\pm0.015\times10^{-10}$ erg\,s$^{-1}$\,cm$^{-2}$. Errors on the bolometric flux account for random and correlated (e.g., slope errors in the flux calibration) errors in the combined spectrum, as well as measurement and zero point errors in the photometry, as detailed in \citet{Mann2015b}. When combined with the kinematic distance, this yields a luminosity of $8.4\pm1.4\times10^{-3}L_{\odot}$. 

{\it Effective temperature}: We derived an effective temperature (\teff) from our optical spectrum following the procedure from \citet{Mann2013}. To briefly summarize, \citet{Mann2013} compare optical spectra of M dwarfs with BT-SETTL CIFIST models\footnote{\href{https://phoenix.ens-lyon.fr/Grids/BT-Settl/CIFIST2011}{https://phoenix.ens-lyon.fr/Grids/BT-Settl/CIFIST2011}} \citep{Allard2011}. By masking out regions of the spectrum that are poorly reproduced by the models, \citet{Mann2013} reproduced the \teff\ scale from long-baseline interferometry \citep{Boyajian2012} to 60\,K, which we adopted as the error on our measurement. This procedure yields a \teff\ of 3180$\pm$60\,K. 

{\it Mass, radius and density}: To estimate the stellar mass, radius, and density, we used the \citet{Mann2015b} relations between absolute $K_s$-band magnitude ($M_{K_s}$) and metallicity and stellar radius/mass. The radius relation was calibrated using angular diameter measurements from long-baseline optical interferometry \citep{Boyajian2012}. The mass relation is slightly model dependent, but reproduces the empirical mass-luminosity relation from \citet{Delfosse2000}, and in combination the mass and radius relations reproduce the mass-radius relation from low-mass eclipsing binaries \citep{Feiden2012a, Mann2015b} within errors. Accounting for errors in the distance, [Fe/H], $K_s$ magnitude, and scatter in the relations, we derived a radius of $0.295\pm0.020\,R_\odot$ and a mass of $0.294\pm0.021\,M_\odot$. For the stellar density we also considered that errors are correlated, i.e., if the distance is greater, both the mass and radius increase together. Accounting for this via Monte Carlo simulation, we find a density of $11.3^{+1.7}_{-1.5}\rho_\odot$.  

\citet{Mann2015b} relations are primarily based on stars older than 1\,Gyr, while \name\ is relatively young (650--800 Myr). However, at this age \name\ is expected to be on the main sequence, and the \citet{Mann2015b} relations include some stars with similar activity levels (as measured by H$\alpha$) as \name. As a check we also derived the stellar radius using the Stefan--Boltzmann law and our above luminosity and \teff. This yields a radius of $0.301\pm0.032\,R_\odot$, $<1\sigma$ from our estimate above. We note that this method is not completely independent of the $M_{K_s}$-$R_*$ relation, as they both rely on the same distance and $K_s$ measurement.

\begin{figure}
	\centering
	\includegraphics[width=0.95\columnwidth]{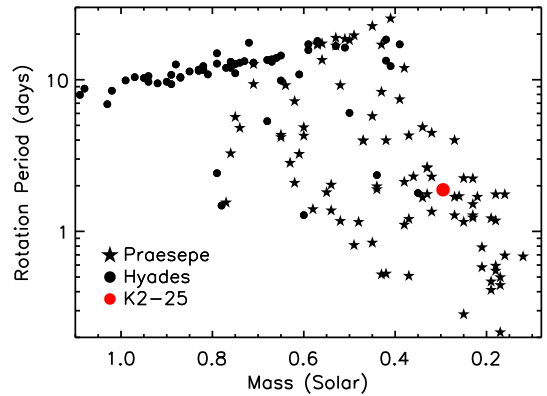} 
	\caption{Rotation period as a function of stellar mass for stars in the Hyades and Praesepe clusters taken from \citet{2011ApJ...740..110A}, \citet{2011MNRAS.413.2595S}, and \citet{2011MNRAS.413.2218D}. Praesepe targets are similar in age to that of the Hyades (both 650-800\,Myr) and have more rotation period measurements at low masses in the literature. \name\ is shown in red. Errors on mass measurements are typically $\simeq$10\%. }
	\label{fig:prot}
\end{figure}

{\it Rotation period}: The light curve from \citet{Vanderburg2014} shows $\sim1\%$ amplitude periodic variation due to rotation and spot coverage. We calculated the autocorrelation function of the light curve and found a peak at  $P=1.881\pm0.021$ days, with a harmonic at 0.940$\pm$0.005 days; we consider the former the rotation period of the star.  Errors on the rotation period were determined by fitting the autocorrelation function around the peak as a Gaussian and do not consider (potential) sources of systematic error such as aliasing, differential rotation, or short-lived spots \citep[e.g.,][]{Aigrain2015}. However, periodicity does not change over the 71-day observing period from {\it K2} despite changes in the amplitude. Further, this rotation period is consistent with stars of similar mass and age in the Hyades and Praesepe clusters (Figure~\ref{fig:prot}).

{\it \vsini\footnote{$i_*$ is used for the stellar sky-projected inclination to distinguish it from $i$, the inclination of the planet's orbit.}}: We estimated the level of rotational broadening in \name\ from our stacked IGRINS spectrum. We compared our IGRINS spectrum with a BT-SETTL model with a \teff, log\,$g$, and [M/H] of 3200, 5.0, and 0.0 (roughly consistent with our calculations above), which we artificially broaden with the IDL code \textit{lsf\_rotate} \citep{1992oasp.book.....G,2011ascl.soft09022H}. Each of the IGRINS orders is fit separately, excluding those for which the S/N is too low ($<20$). We normalized each order and the appropriate region of the model with a 150-pixel (much larger than a given feature) running median after masking out regions of strong ($>30\%$) telluric absorption. To account for instrumental broadening, we simultaneously fit the telluric lines in each order, which we extract from the A0V star spectrum (see Section~\ref{sec:igrins}). We assume that instrumental broadening is Gaussian and that the telluric lines have negligible intrinsic width. The instrumental broadening typically has a full-width half-max of 0.3-0.5\,\AA, consistent with the resolution of the spectrograph. We applied the broadening derived for each order to the model. We then fit the model to the spectrum, letting \vsini\ float from 0 to 50\,\kms, calculating $\chi^2$ at each step. We adopted the median and standard deviation of the \vsini\ measurements across all fit orders as the final measurement and error, which was $7.8\pm0.5$\,\kms.

{\it Sky projected inclination}: The combination of our \vsini, rotation period, and stellar radius enables us to calculate the (sky-projected) rotational inclination ($i_*$) of \name. We first calculate the equatorial velocity ($V_{\rm{eq}}$):
\begin{equation}
V_{\rm{eq}} = \frac{2\pi R_*}{P_{\rm{rot}}},
\end{equation}
where $P_{\rm{rot}}$ is the stellar rotation period. This yields a velocity of $7.9\pm0.5$\kms. We follow the formalism from \citet{Morton2014b} to convert \vsini\ and $V_{\rm{eq}}$ to a posterior in $\cos(i_*)$, which handles regions of the posterior where \vsini$>V_{\rm{eq}}$ (which is physically impossible). The resulting posterior gives a lower limit on stellar inclination of $i_*>52^{\circ}$ at 99.7\% confidence (3$\sigma$), and $i_*>72^{\circ}$ at 68.3\% (1$\sigma$). 

All derived stellar parameters for \name\ are listed in Table~\ref{tab:params}.

\begin{deluxetable}{l c l }
\tabletypesize{\scriptsize}
\tablecaption{Parameters of \name \label{tab:params}}
\tablewidth{0pt}
\tablehead{
\colhead{Parameter} & \colhead{Value} & \colhead{Source}
}
\startdata
\multicolumn{3}{c}{\hspace{1cm}Astrometry} \\
$\alpha$ R.A. (hh:mm:ss) & 04:13:05.61 & EPIC\\
$\delta$ Dec. (dd:mm:ss) & +15:14:52.00 & EPIC\\
$\mu_{\alpha}$ (mas~yr$^{-1}$) & 120.6 $\pm$ 3.3 & \cite{Roser2011} \\
$\mu{\delta}$ (mas~yr$^{-1}$) & $-$21.1 $\pm$ 3.2 & \cite{Roser2011} \\
\hline
\multicolumn {3}{c}{\hspace{1cm}Photometry} \\
$B$ (mag) & 17.760 $\pm$ 0.289 & APASS \\
$V$ (mag) & 15.881 $\pm$ 0.030 & APASS \\
$g$ (mag) & 16.730 $\pm$ 0.010 & SDSS \\
$r$ (mag) & 15.235 $\pm$ 0.031 & SDSS \\
$i$ (mag) & 13.760 $\pm$ 0.010 & SDSS \\
$z$ (mag) & 12.820 $\pm$ 0.010 & SDSS \\
$J$ (mag) & 11.303 $\pm$ 0.021 & 2MASS\\
$H$ (mag) & 10.732 $\pm$ 0.020 & 2MASS\\
$K_s$ (mag) & 10.444 $\pm$ 0.019 & 2MASS\\
$W1$ (mag) & 8.443 $\pm$ 0.023 & $WISE$ \\
$W2$ (mag) & 8.424 $\pm$ 0.021 & $WISE$\\
$W3$ (mag) & 8.322 $\pm$ 0.055 & $WISE$ \\
\hline
\multicolumn{3}{c}{\hspace{1cm}Derived Properties} \\
Rotation period (days) &   1.88 $\pm$ 0.02 & This paper \\
Barycentric RV (\kms) & 38.64 $\pm$ 0.15 & This paper \\
$U$ (\kms) & $-42.4\pm  1.2$ & This paper \\
$V$ (\kms) & $-18.4\pm  3.2$ & This paper \\
$W$ (\kms) & $ -1.8\pm  2.4$ & This paper \\
$X$ (pc) & $-39.8\pm  6.3$ & This paper \\
$Y$ (pc) & $  +1.2\pm  0.2$ & This paper \\
$Z$ (pc) & $-18.8\pm  3.0$ & This paper \\
Distance (pc) &   $45.7 \pm 3.3$ & This paper\tablenotemark{a}\\
EW (H$\alpha$) (\AA) & $-$3.1 $\pm$ 0.1 & This paper \\
\vsini\ (km~s$^{-1}$) & $7.8\pm0.5$  & This paper \\
$i_*$ (degrees) & $>72$ & This paper \\
Spectral type &  M4.5 $\pm$ 0.3 & This paper\\
$[$Fe/H$]$ & 0.15 $\pm$ 0.03 & This paper\tablenotemark{b}\\
\teff\ (K) & 3180 $\pm$ 60 & This paper\\
$M_*$ ($M_\odot$) & 0.294 $\pm$ 0.021 & This paper \\
$R_*$ ($R_\odot$) & 0.295 $\pm$ 0.020& This paper \\
$L_*$ ($L_\odot$) & ($8.4\pm1.4)\times10^{-3}$ & This paper \\
$\rho_*$ ($\rho_\odot$) &  $11.3^{+1.7}_{-1.5}$ & This paper
\enddata
\tablenotetext{a}{The distanced derived from cluster membership and kinematics of \pname\ (see Section~\ref{sec:params}).}
\tablenotetext{b}{This is the weighted mean of our own measurements from SpeX and literature measurements for the Hyades cluster (see Section~\ref{sec:params}).}
\end{deluxetable}

\section{{\it K2} light curve Analysis}\label{sec:transit} 
For our analysis we relied on the light curve from \citet{Vanderburg2014}, which used a $3\times3$ pixel aperture centered on \name. We also downloaded and repeated our analysis using the Pre-search Data Conditioning \citep[PDCSAP;][]{Stumpe2012} light curve released by the \kepler{} and {\it K2} Science Center. Both light curves are shown in Figure~\ref{fig:lc}. Our final results were consistent using either light curve, although the residuals of our transit analysis were smaller when using the reprocessed \citet{Vanderburg2014} light curve, so we report values from that analysis. 

\pname\ was initially identified by eye in the PDCSAP light curve, as the periodic 1\% dips (Figure~\ref{fig:lc}) can be seen even through the strong rotational variability. 

\subsection{Search for additional transiting planets}\label{sec:idenfitication}

We performed a systematic search for other transit signals in the {\it K2} light curve of \name. We started with the light curve from \citet{Vanderburg2014} and applied several additional corrections to filter stellar rotational variability and flaring. First, a power spectrum was generated using the Lomb-Scargle algorithm \citep{Scargle1981} and significant, isolated peaks were identified with false-positive probability (FAP) $<0.01$. These frequencies were filtered from the light curve and a running median with a 1-day window was also removed (the window is much larger than the expected duration of any transit). A robust standard deviation was calculated using the algorithm of \citet{Tukey1977} and $>3\sigma$ positive excursions were replaced with median values. This data set was then searched for periodic transit-like signals using the box-least-squares algorithm from \citet{Kovacs2002}. A second-order trend in the power spectrum was removed, and signals with FAP $<0.01$ were identified by calculating the Signal Detection Efficiency (Eqn. 6 in \citet{Kovacs2002}) and evaluating the significance using the cumulative Gaussian with the parameters set to the values found by \citet{Kovacs2002} for pure noise. These candidate signals were then analyzed in detail and their S/N calculated. Besides the 3.485-day signal investigated here, no other signals with periods between 0.3 and 20 days were detected.

\subsection{Simultaneous-fit {\it K2} light curve}\label{sec:simultfit}

\begin{figure*}
	\centering
	\includegraphics[width=0.95\textwidth]{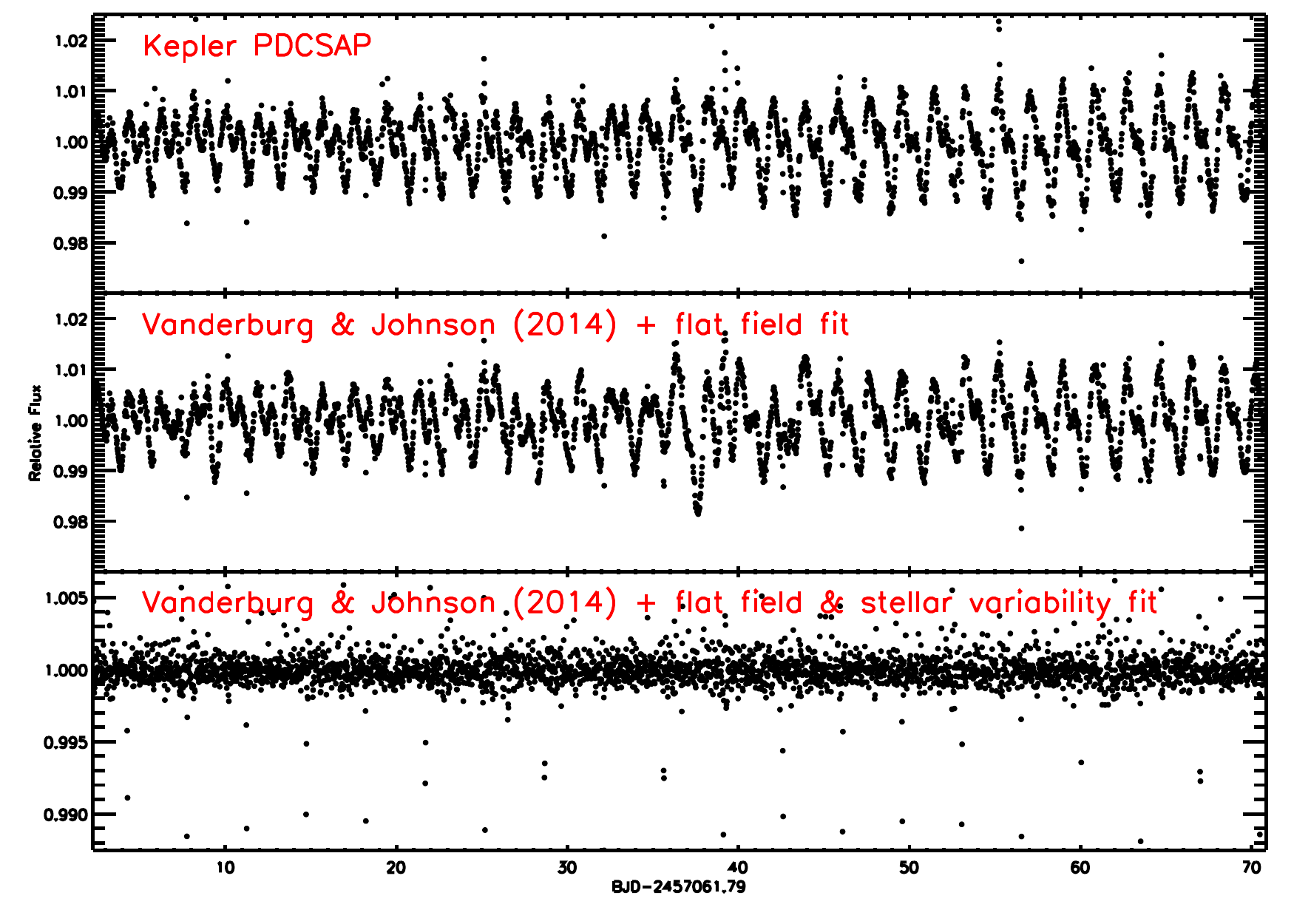} 
	\caption{light curve of \name\ taken by the {\it K2} spacecraft. The top panel is the PDCSAP flux provided by the \kepler{} and {\it K2} Science Center.  The middle panel is the  \citet{Vanderburg2014} light curve with simultaneous fitting of the {\it K2} flat field, while the bottom panel shows the \citet{Vanderburg2014} light curve with simultaneous fitting of the flat field and stellar variability (see Section~\ref{sec:simultfit}). All light curves have been normalized to 1 and the time zeroed to the start of the {\it K2} observations. Some data points ($<1\%$) that we attribute to flares are off the top of the panels. The shape and features seen in the light curves are insensitive to choice of aperture size, as the region is free of significant contaminating flux from background stars (see Figure~\ref{fig:poss}).}
	\label{fig:lc}
\end{figure*}

The \citet{Vanderburg2014} pipeline is optimized for producing light curves of slowly rotating stars, and \name's rapid high amplitude photometric variability resulted in uncorrected systematic effects in the original \citet{Vanderburg2014} light curve. Once we identified the transit we reprocessed the {\it K2} light curve using the same procedure as \citet{Becker2015}, simultaneously fitting for the stellar activity signal, the {\it K2} flat field, and the transits of \pname\ using a Levenberg-Marquardt minimization algorithm \citep{Markwart2009}. We modeled the stellar variability as a spline with breakpoints every 0.2 days, the {\it K2} flat field as splines in {\it K2} image centroid position with breakpoints roughly every 0.4 arcseconds, and the transits with a \citet{MandelAgol2002} model, while taking into account the 30-minute long-cadence integration time. This processing effectively removed the systematics from the {\it K2} pointing jitter and resulted in improved photometric precision. We show both light curves as well as the PDCSAP light curve in Figure~\ref{fig:lc}.

\subsection{Transit Fitting}\label{sec:transitfit}

We fit the flattened light curve with a Monte Carlo Markov Chain (MCMC) by utilizing the {\textit e}mcee Python module \citep{Foreman-Mackey2013} and the \textit{batman} tool \citep{Kreidberg2015}, which utilizes the \citet{MandelAgol2002} transit model. We oversampled and binned the model to the \kepler{} cadence to handle light-curve distortion from long integration times \citep[see][for a discussion of this issue]{Kipping:2010lr}. We sampled the planet-to-star radius ratio ($R_P/R_*$), impact parameter ($b$), orbital period ($P$), epoch of the first transit midpoint ($T_0$), two parameters that describe the eccentricity and argument of periastron ($\sqrt{e}\sin(\omega)$ and $\sqrt{e}\cos(\omega)$), bulk stellar density ($\rho_*$), and two limb-darkening parameters ($q1$ and $q2$). We assumed a quadratic limb-darkening law and use the triangular sampling method of \citet{Kipping2013} in order to uniformly sample the physically allowed region of parameter space. MCMC chains were run using 100 walkers, each with 200,000 steps after a burn-in phase of 10,000 steps. 

The transit duration is uniquely determined from the other fitted transit parameters following the formulae from \citet{Seager:2003lr}, but using $\rho_*$ as the free parameter rather than transit duration enables us to apply a prior on $\rho_*$ using our values derived in Section~\ref{sec:params}. This in turn let us constrain $e$ and $\omega$, which are generally hard to measure owing to their minimal impact on the observed light curve \citep[similar methods are used in][]{Dawson:2012fk,2012MNRAS.421.1166K}. As explained in \citet{Van-Eylen2015}, directly exploring $e$ and $\omega$ biases the eccentricity to higher values owing to the cutoff at zero, while sampling uniformly in $\sqrt{e}\sin(\omega)$ and $\sqrt{e}\cos(\omega)$ between -1 and 1 is unbiased and still provides uniform sampling in $e$ and $\omega$ \citep[see][for more detailed discussions of this issue]{1971AJ.....76..544L,Ford2006, Anderson2011, 2013PASP..125...83E}.

We applied a prior drawn from the model-derived limb-darkening coefficients from \citet{Claret2011} assuming a quadratic limb-darkening law. We interpolate our stellar parameters (log~$g$, \teff, [Fe/H]; see Section~\ref{sec:params}) onto the \citet{Claret2011} grid of limb-darkening coefficients from the PHOENIX models, accounting for errors from the finite grid spacing, errors in stellar parameters, and variations from the method used to derive the coefficient (Least-Square or Flux Conservation). This yielded priors of $0.45\pm0.1$ and $0.35\pm0.1$ for the linear and quadratic limb-darkening coefficients, respectively, which we propagated to $q_1$ and $q_2$ using the formulae in \cite{Kipping2013}. We uniformly sampled the impact parameter over -1.2 to +1.2 (to allow for grazing transits), orbital period over 0 to 100 days, and mid-time of the first transit over $\pm$1.5 days (about half the period) from the value identified in our L-S analysis (Section~\ref{sec:idenfitication}). 

\begin{figure}
	\centering
	\includegraphics[width=0.95\columnwidth]{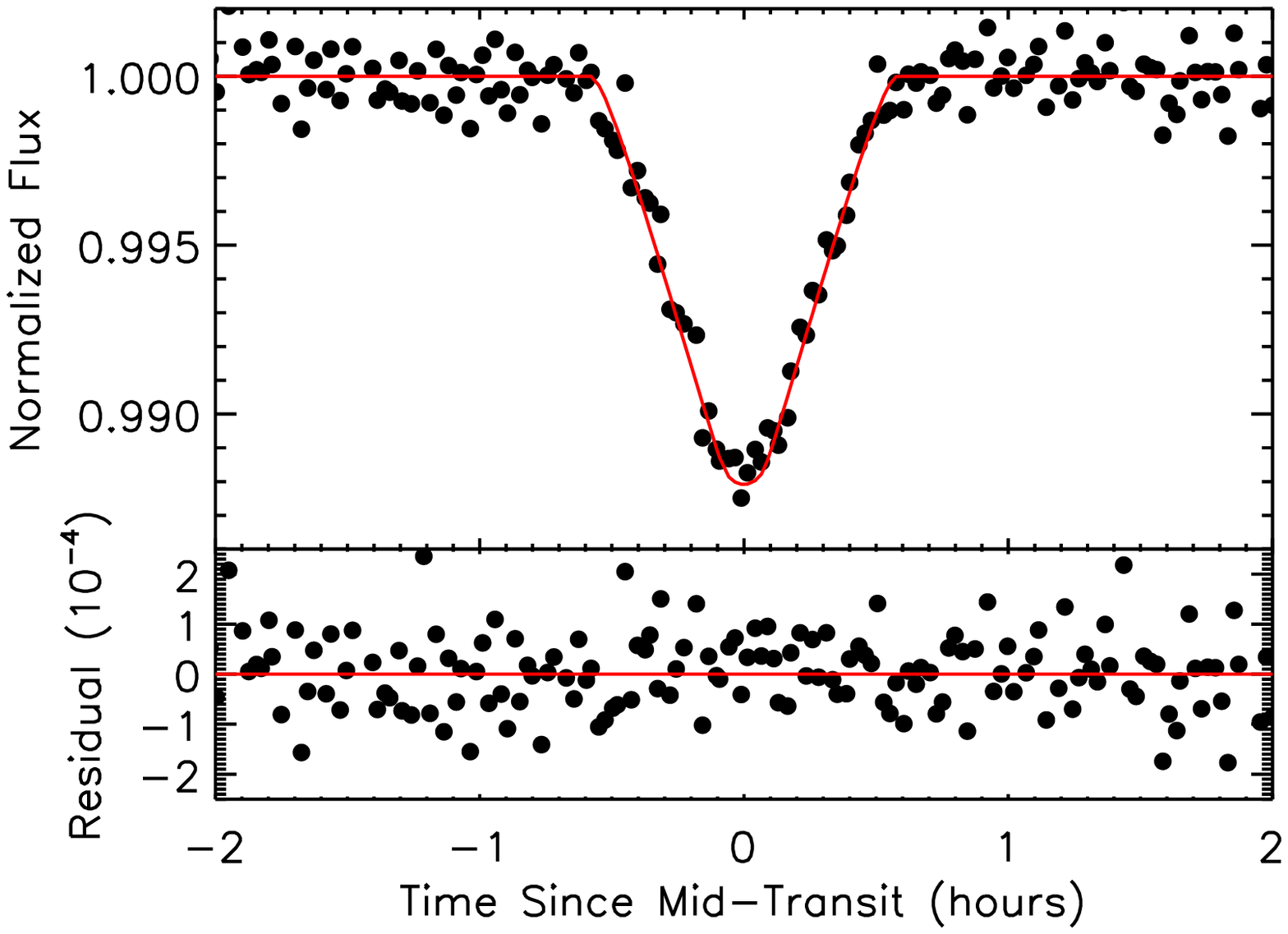} 
	\caption{Phase-folded light curve of \name's transit (black points). The red line shows the best-fit (highest likelihood) model from our MCMC fit (Section~\ref{sec:transit}). The bottom panel shows the fit residuals. }
	\label{fig:transitfit}
\end{figure}

\begin{figure*}
	\centering
	\includegraphics[width=0.46\textwidth]{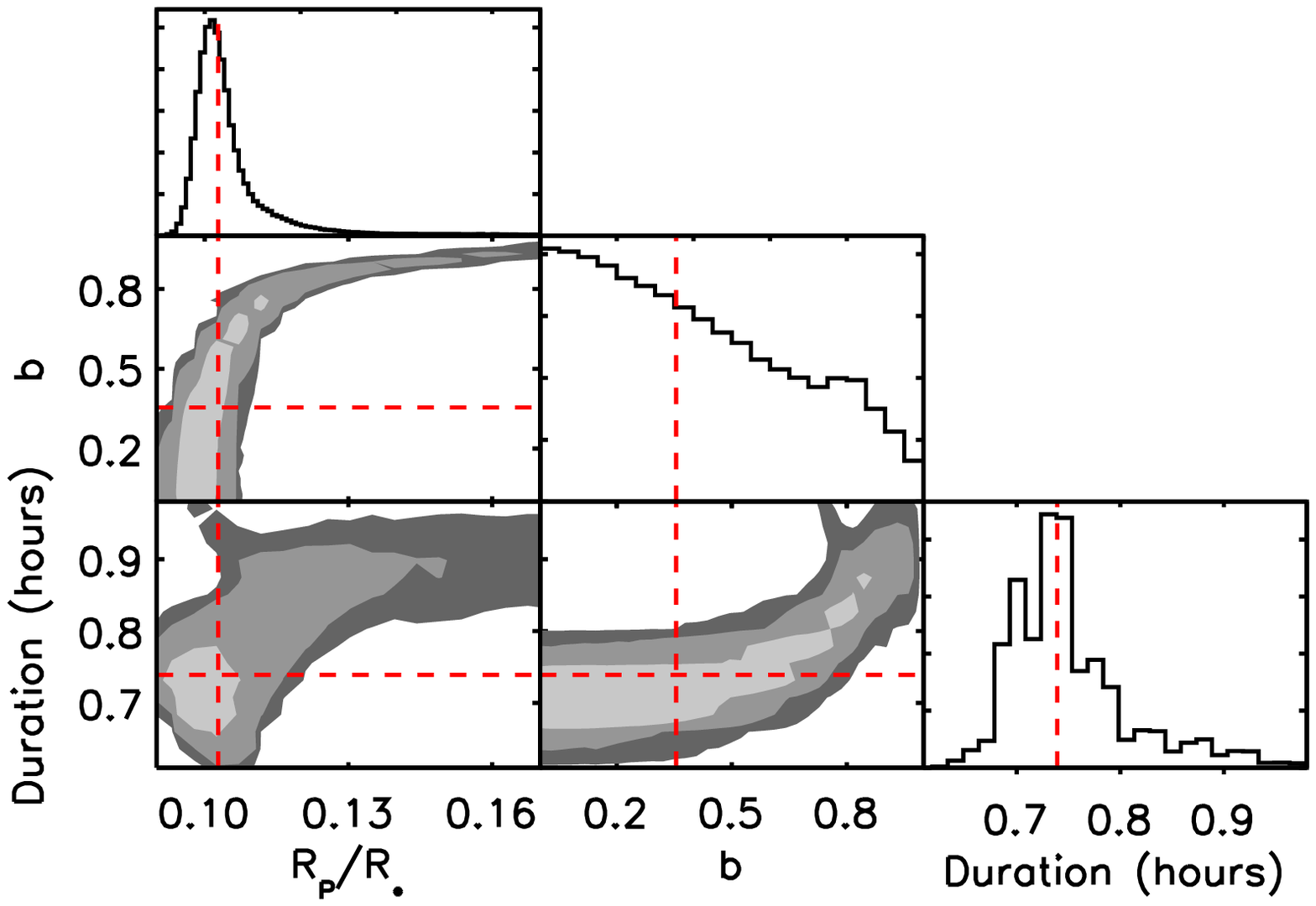} 
	\includegraphics[width=0.46\textwidth]{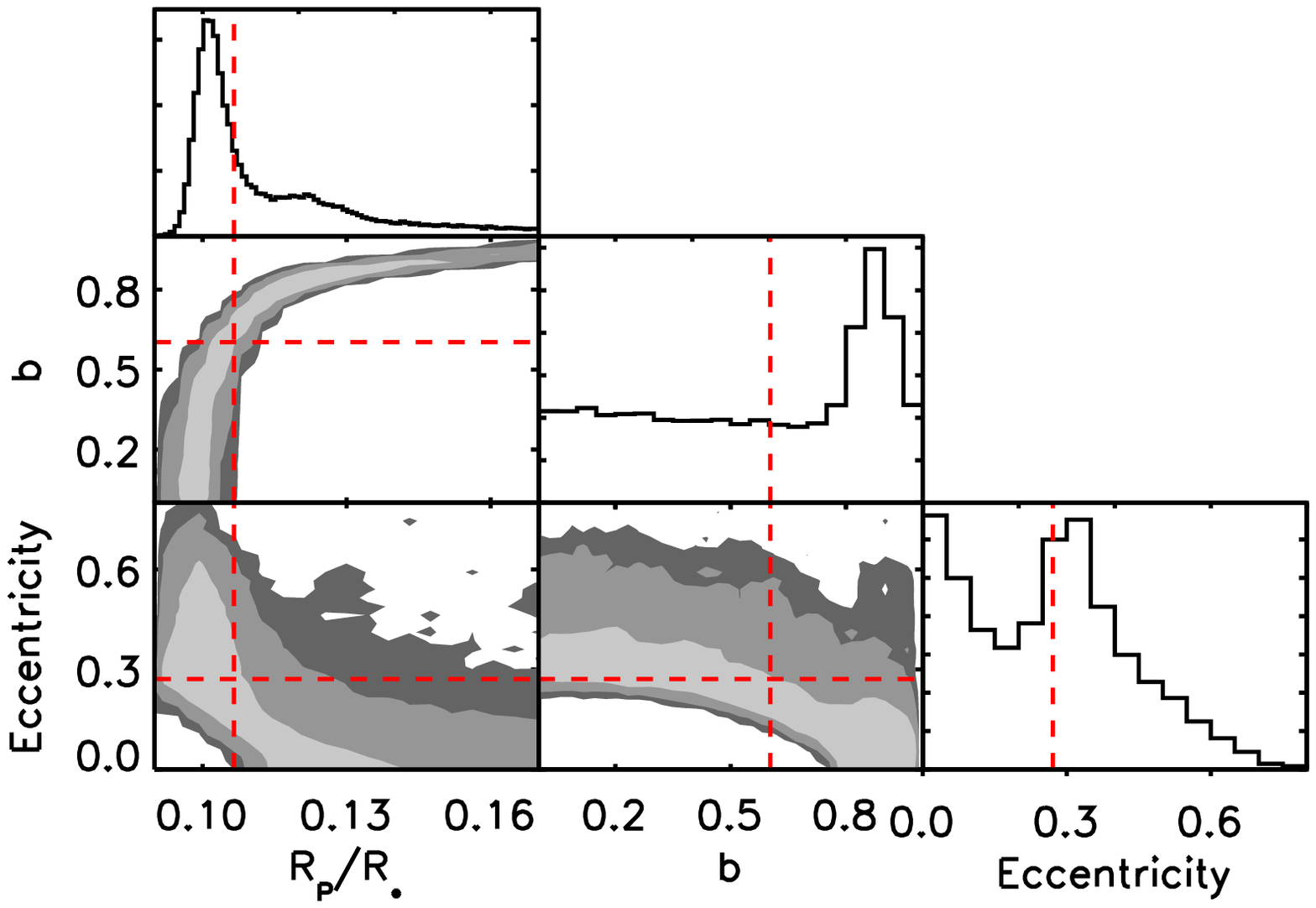} 
	\caption{Posteriors from our MCMC fit. Median values for each parameter are marked with red dashed lines. Gray scaling corresponds to 1$\sigma$, 2$\sigma$, and 3$\sigma$ (from light to dark). The left panel shows the fit with eccentricity and argument of periastron fixed at 0 and no prior on density, while the right panel shows a fit with $\sqrt{e}sin(\omega)$ and $\sqrt{e}cos(\omega)$ allowed to float and a prior on stellar density from our analysis in Section~\ref{sec:params}. For the latter fit we show the eccentricity posterior, but since eccentricity is fixed in the former, we instead show the transit duration posterior. }
	\label{fig:transitparams}
\end{figure*}

We fit the light curve twice: once with $\sqrt{e}\sin(\omega)$ and $\sqrt{e}\cos(\omega)$ fixed at 0 and no prior on $\rho_*$, and once with $\sqrt{e}\sin(\omega)$ and $\sqrt{e}\cos(\omega)$ limited to $-1$ to $+1$ and under uniform priors, and with a prior on $\rho_*$ using our values derived in Section~\ref{sec:params}. We report the results of both transit fits in Table~\ref{tab:planet}. For each parameter we report the median value with the `errors' as the 84.1 and 15.9 percentile values (corresponding to 1$\sigma$ for Gaussian distributions). The model light curve with the best-fit parameters (highest likelihood) is shown for the latter fit in Figure~\ref{fig:transitfit} alongside the {\it K2} data. We also show posteriors and correlations for a subset of parameters in Figure~\ref{fig:transitparams}.

\renewcommand{\arraystretch}{1.5}
\begin{deluxetable*}{l l l l l l l l l l l }
\tablecaption{Transit Fit Parameters}
\tablewidth{0pt}
\tablehead{
\colhead{Parameter} & \colhead{Fit 1\tablenotemark{a}} & \colhead{Fit 2\tablenotemark{a} (Preferred)} 
} 
\startdata
Period (days) & $3.484552^{+0.000036}_{-0.000044}$ & $3.484552^{+0.000031}_{-0.000037}$ \\
$R_P/R_*$ & $0.1028^{+0.0080}_{-0.0037}$ & $0.1065^{+0.0286}_{-0.0065}$ \\
T$_0$ (BJD\tablenotemark{b}-2400000) & $57062.57935^{+0.00049}_{-0.00024}$ & $57062.57935^{+0.00049}_{-0.00024}$ \\
Impact parameter & $0.35^{+0.36}_{-0.25}$ & $0.60^{+0.29}_{-0.42}$ \\
Duration\tablenotemark{c} (hours) & $0.74^{+0.06}_{-0.04}$ & $0.79^{+0.09}_{-0.17}$ \\
Inclination\tablenotemark{c} (degrees) & $89.5^{+0.4}_{-0.9}$ & $88.3^{+1.2}_{-0.7}$ \\
Eccentricity & 0 (fixed) & $0.27^{+0.16}_{-0.21}$ \\
$\omega$ (degrees) & 0 (fixed) &  $62^{+44}_{-39}$ \\
\hline
$R_P$\tablenotemark{d} ($R_\earth$) & $ 3.31^{+0.34}_{-0.25}$\,R$_\earth$ & \prad 
\enddata
\tablenotetext{a}{Fit 1 is done with $e$ and $\omega$ fixed at 0 and a uniform prior on $\rho_*$, while fit 2 is done with $\sqrt{e}sin(\omega)$ and $\sqrt{e}cos(\omega)$ limited to $-1$ to $+1$ under uniform priors and with a prior on $\rho_*$ from our analysis in Section~\ref{sec:params}.}
\tablenotetext{b}{ BJD is given in Barycentric Dynamical Time (TBD) format.}
\tablenotetext{c}{ For both fits stellar density and impact parameter are the fitted parameters (instead of transit duration and orbital inclination). We report the duration and inclination derived from the other fit parameters here for convenience. }
\tablenotetext{d}{Planet radius is derived using our stellar radius from Section~\ref{sec:params}.}
\label{tab:planet}
\end{deluxetable*}

Both fits have a noticeable tail in the $R_P/R_*$ distribution owing to poor sampling of the transit duration and a degeneracy with impact parameter. The transit duration is only slightly longer than the integration time, so it is difficult to completely rule out partially grazing ($b>0.9$) orbital solutions. The MCMC is able to achieve a reasonable fit to the data with large $R_P/R_*$ values by simultaneously adjusting the impact parameter and transit duration. However, such solutions are also disfavored compared to solutions with a lower impact parameter and shorter transit duration (although less so for the fit with the prior on $\rho_*$). This issue could be further complicated if model limb-darkening parameters turn out to be systematically erroneous for cool stars, although fits of high-quality light curves suggest that the model parameters are at least roughly correct \citep[e.g.,][]{2014Natur.505...69K, Swift2015}. We note that follow-up observations from the ground at higher cadence could significantly mitigate this degeneracy by resolving the transit duration, particularly NIR observations, where limb-darkening has a smaller impact.

The two different MCMC fits give consistent values for all fit values. The main difference is that the fit with the $\rho_*$ prior favors a higher impact parameter and hence has power at large radii. The $\rho_*$-prior fit is consistent with $e=0$, although a modest value ($e\simeq0.3$) is slightly favored. Again, better data to resolve the transit duration would help significantly here, as $e$ is highly degenerate with $b$ and $\tau$ when using a prior on $\rho_*$. We adopt the second fit for our planet parameters as the additional constraints on $\rho_*$ should provide a more accurate picture of the true planet parameters. Further, \pname\ has a tidal circularization timescale of $\sim1$\,Gyr \citep{Goldreich1966} and hence probably has not lost its initial eccentricity. Combined with our stellar radius from Section~\ref{sec:params} the latter fit yields a planet radius of \prad.

\section{False Positive Analysis}\label{sec:fpp}

\subsection{Background Star}\label{sec.background}

\begin{figure}
	\centering
	\includegraphics[width=0.46\textwidth]{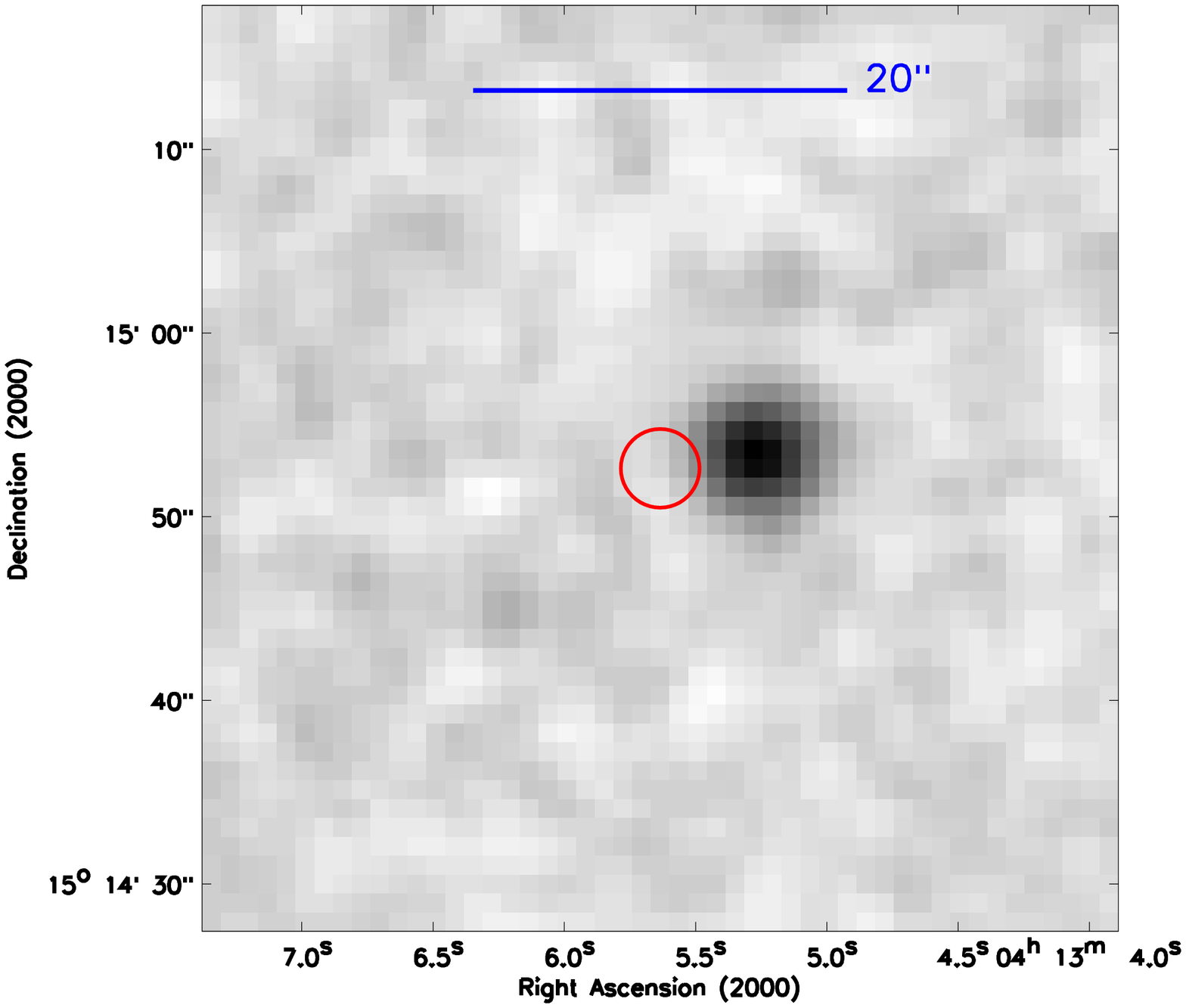} 
	\includegraphics[width=0.46\textwidth]{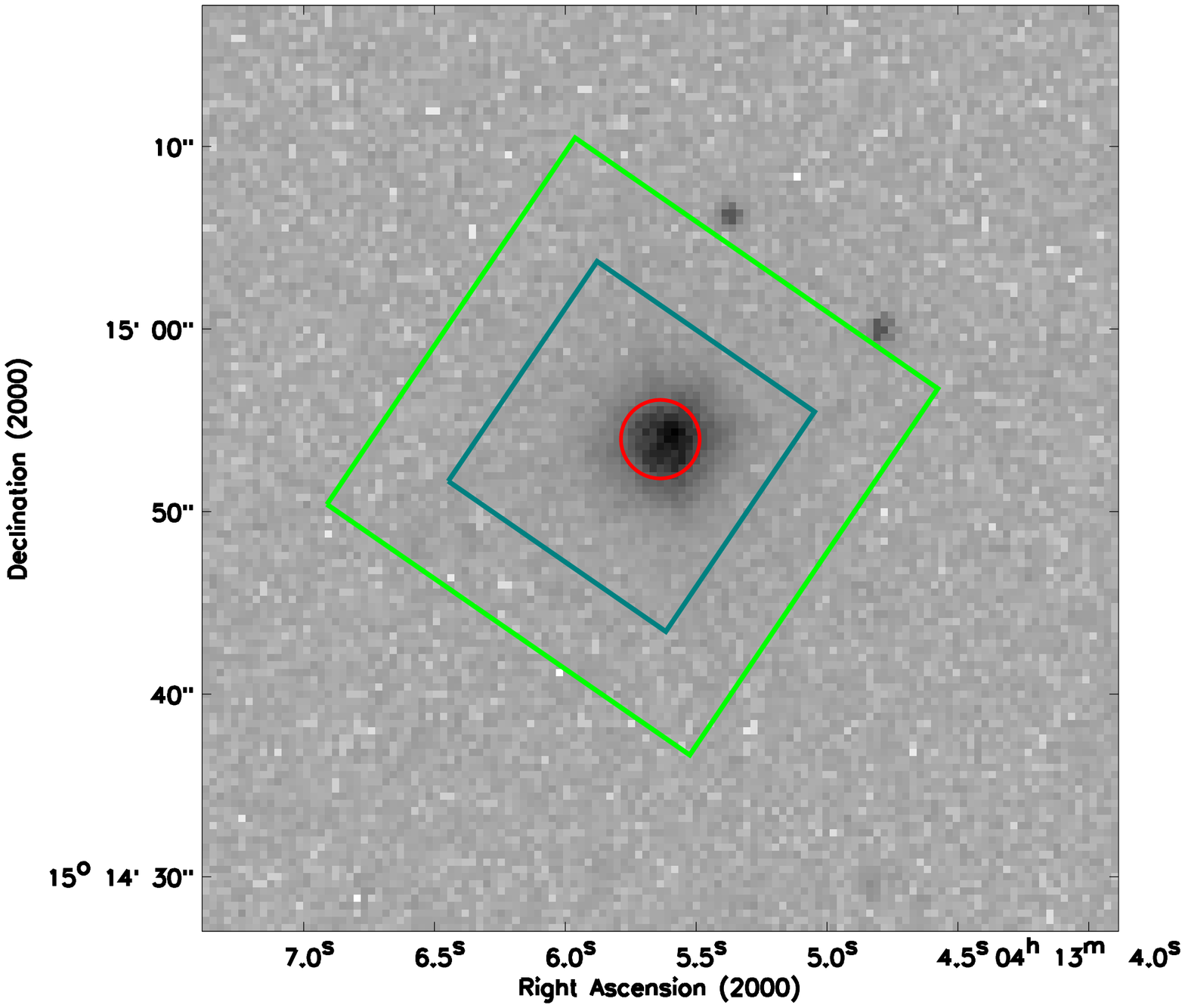} 
	\caption{Archive images of \name\ from DSS (top) and SDSS (bottom). Both images have the same scale, and a 20\arcsec\ bar is shown in blue on the DSS image. The current location of the object is shown as a red circle in both images. Because the DSS image is from 1953, the target was $\sim$7\arcsec\ from its current position, revealing potential unresolved background stars. Boxes corresponding to {\it K2} apertures (5$\times$5 and 3$\times$3 {\it K2} pixels) are shown in green and teal in the SDSS image. We use the smaller (9 pixel) aperture to cut out the faint background star. Because of the large {\it K2} PSF, two background stars visible in SDSS still contaminate the smaller aperture, but both of these stars are too faint to reproduce the 1\% transit depth. }
	\label{fig:poss}
\end{figure}

We calculated a posterior probability that an unrelated background star could be responsible for the transit signal, i.e., as an eclipsing binary (EB). We followed the procedure described in detail in \citet{Gaidos2016}, which is summarized here. The Bayesian probability was calculated with a prior based on a model of the background stellar population and a likelihood calculated from observational constraints, i.e., (1) a background star has to be bright enough to produce a transit signal given a maximum possible eclipse depth of 50\%, (2) the density of the star must be consistent with the transit duration; (3) the star is not visible in the 1953 POSS red and blue image in which \name, owing to its proper motion over 6 decades, is displaced by about 7\arcsec\ (Figure~\ref{fig:poss}); and (4) the star is not visible in our NIRC2 AO imaging (Section~\ref{sec:ao}).

We used a model stellar population calculated by TRILEGAL version 1.6 \citep{Vanhollebeke2009}. We calculated the stellar population at the position of \name\ to $K_p = 22$ equivalent over a field of 10 square degrees (to reduce counting noise). The faint limit is several magnitudes deeper than the faintest EB diluted that could possibly produce the transit signal after dilution by \name\ ($K_p = 14.53$\,mag). The false positive probability (FPP) was calculated by the method of Monte Carlo; model stars were placed at random locations in a circular field 16\arcsec\ (4 \kepler\ pixels) in radius centered on \name. Stars were discarded or retained based on the contrast criterion $\Delta K_p < - 2.5 \log (\delta / R)$, where $\delta = 0.012$ is the transit depth and $R$ is a pixel response function interpolated from the \kepler\ Instrument Handbook supplement values for the appropriate detector channel (13). We also discarded stars brighter than $K_p = 19$\,mag based on the DSS POSS 1 image and stars with a contrast in the infrared $K$ band brighter than the $7\sigma$ detection limit in our Keck 2-NIR2-AO imaging. We weighted each remaining star with the probability that a transit of an object with the observed orbital period (3.485~d) would have the observed duration ($\simeq$47~minutes) if placed around the background star, versus being placed around \name. This calculation requires a probability distribution for $e$, and we adopted a Rayleigh distribution with mean $e = 0.1$, appropriate for short-period binaries. We then summed up the number of stars in the 16'' circle and divided by the ratio of the circular solid angle to 10 square degrees. We find a  FPP of $\approx 4.5 \times 10^{-8}$, a consequence of the depth of the transit, brightness of \name, and the low background star counts at this moderate Galactic latitude. 

\subsection{Eclipsing Companion to \name}

Because of the short orbital period, small stellar size, and long observing cadence, a V-shaped transit is expected for a low-eccentricity orbit, even if the eclipse/transit is partial. However, the V shape we see also leaves open the possibility that the observed transit is actually a grazing eclipse from a stellar companion with a 3.485-day orbital period. This is reflected in the tail in the posterior of the $R_P/R_*$ distribution. Although the posterior never passes into values consistent with stars, the fits assume that the eclipsing body is not  luminous, and hence this alone does not rule out the possibility that the transit is due to a star. Further, although no secondary eclipse is seen and the even and odd transits have consistent depths, there are regions of inclination, $e$, and $\omega$ space where there would be no secondary eclipse. 

Instead, we can rule out this possibility using the RVs derived from our IGRINS spectra (see Section~\ref{sec:igrins}, and Section~\ref{sec:hyades}). In theory the relative RVs from our IGRINS data should be more precise than the absolute velocities, but this does not account for astrophysical jitter common to young stars. Empirical measurements of Hyades-age Sun-like stars in the optical suggest variations of 50\mps\ from activity \citep{2004AJ....127.3579P,2015csss...18..759H}. Based on the amplitude of spot variations we see in the {\it K2} light curve and the \vsini\ measured from our IGRINS spectrum, the spot-induced RV jitter should be $\sim$120\mps\ in the \kepler{} bandpass, but observations of T Tauri stars suggest that this is smaller by a factor of two to three at the wavelength of our IGRINS data \citep[2.2\um;][]{Crockett2012}. 

\begin{figure}
	\centering
	\includegraphics[width=0.95\columnwidth]{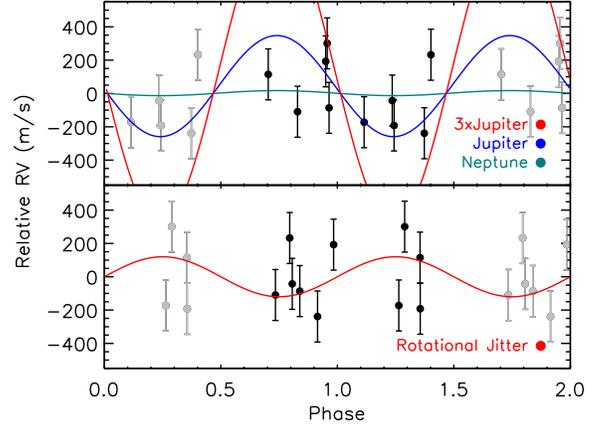} 
	\caption{RVs from IGRINS phased to the planet's orbital period (top, 3.485\,days) and star's rotation period (bottom, 1.88 days). Two phases are shown, but repeat measurements are shown in gray. In the top panel the expected signal from a Neptune-mass, Jupiter-mass, and 3$\times$Jupiter mass planet (with $e=0$ and $P=3.485$\,days) are shown as teal, blue, and red lines, respectively. In the bottom panel we show the predicted jitter from the $\sim$1.5\% spot-induced variations in the {\it K2} light curve and the \vsini\ measurement of $\sim$8\,\kms, although this is expected to be smaller in the $K$ band. }
	\label{fig:rv}
\end{figure}

Another complication is the stability of IGRINS for precision RVs. The use of telluric lines to correct the wavelength scale should reduce instrumental variability to 5-20\mps\ \citep[e.g.,][]{2010A&A...515A.106F,Blake2010}, but correlated errors may persist and radial velocity stability should be tested empirically. So instead of the expected errors of 50\mps\ derived from the scatter in RV variations between orders, we adopt the larger error (typically 150-160\mps) derived from scatter from using different RV templates. We report the resulting velocities and errors for each epoch in Table~\ref{tab:rvs}. We show the RVs in Figure~\ref{fig:rv} phased against both the planet's orbital and the star's rotation period. 

\begin{deluxetable}{l r r}
\tabletypesize{\scriptsize}
\tablecaption{Relative Radial Velocities\label{tab:rvs}}
\tablewidth{0pt}
\tablehead{
\colhead{JD-2,400,000} & \colhead{RV (\mps)\tablenotemark{a}} & \colhead{$\sigma_{\rm{RV}}$ (\mps)}
}
\startdata
 57288.92269 &   300 &   161\\
 57289.89553 &   -42 &   159\\
 57293.85944 &  -238 &   161\\
 57295.86940 &   192 &   158\\
 57319.84005 &  -109 &   158\\
 57320.83534 &  -172 &   159\\
 57321.83510 &   232 &   159\\
 57322.88659 &   114 &   159\\
 57323.79641 &   -85 &   159\\
 57324.76798 &  -191 &   159
\enddata
\tablecomments{\tablenotemark{a}{Radial velocities are quoted with respect to the mean. For the system (absolute) velocity see Table~\ref{tab:params}.}}
\end{deluxetable}

Although the velocity scatter is too large to measure the mass of \pname, we set a limit to the mass on the transiting body by assuming that it is the source of all variation in the RV. To this end we fit the RV data with a simple least-squares minimization \citep{Markwart2009}, fixing $P$ to the value from the transit fit, $e$ to 0, and limiting the eclipse time to a $3\sigma$ range given from the transit fits (but that are otherwise unconstrained) and the inclination to $>80^{\circ}$. We found that the largest mass that is still consistent with the RV values at $5\sigma$ is 3 Jupiter masses, ruling out a grazing EB. Although this alone does not rule out the possibility of a grazing Saturn- or Jupiter-sized planet, such a solution is strongly disfavored by our transit fit (Section~\ref{sec:transit}).

\subsection{Eclipsing Binary Companion to \name}

We considered the possibility that the transit signal is due to a EB bound to \name\ (we consider background EBs in Sec. \ref{sec.background}). To be missed by our AO images any companion must be within $\simeq$10\,AU or be too faint ($\Delta K_p>4$\,mag) to reproduce the transit depth (Figure~\ref{fig:AO}). \citet{Kraus2015} show that planet formation is suppressed by $>85\%$ inward of 60\,AU. We tested the binary companion hypothesis with our Doppler data because the stellar components of a 3.5~d-period system would exhibit significant RV variation  over the baseline of our observations, producing multiple, variable sets of lines in the IGRINS spectra.  We see no second set of lines in any of our NIR spectra, nor is there a second peak in the cross-correlation function.  We determine the significance of this nondetection by simulating ternary (\name\ plus EB companion) systems, adding companions to \name\ drawn randomly from the observed binary mass ratio from \citet{2013ARA&A..51..269D}, but with the companion as a 3.485-day EB with a mass ratio randomly drawn from a uniform distribution. We added BT-SETTL synthetic spectra to our stacked spectrum of \name\ and then searched for a second set of lines or a second peak in the cross-correlation function. We found that 99.8\% of our simulated systems either are insufficiently bright to produce the observed transit depth, would be seen in the AO image, or produce a second peak in the cross-correlation function of one or more of our IGRINS observations. This simulation is also likely an overestimate, as it only simulates triple systems.  

\section{Summary and Discussion}\label{sec:discussion} 

As members of the Hyades have common and well-established ages, metallicities, and distances, their other properties can be constrained more precisely, allowing more rigorous studies of how planets evolve structurally and dynamically with time. M dwarfs are especially interesting targets for exoplanet searches in clusters because their small size facilitates the discovery of smaller planets. To this end we have begun the KELP project to find transiting exoplanets around low-mass cluster members monitored by {\it K2}. \name\ represents the first discovery in our search, as well as the first discovery of a transiting planet in the Hyades. 

We obtained moderate-resolution optical and NIR spectra of \name, used to measure its temperature and luminosity. Multiple epochs of high-resolution, NIR spectra with the IGRINS spectrograph enabled us to confirm \name's membership in the Hyades cluster and rule out the possibility that the transit signal is due to a grazing eclipsing binary. Because the star is a member of the Hyades, we were able to derive a kinematic distance and correspondingly precise stellar parameters. We took advantage of these tight constraints on stellar density to improve the transit fit by applying a prior on $\rho_*$, enabling weak constraints on $e$ and better constraints on $b$. Using RVs from the IGRINS spectra, AO imaging, and the lack of background stars seen in the 1953 POSS data (during which \name\ was in a different location), we are able to confirm the planetary nature of the transit signal. 

Owing to the excellent precision of \kepler, M dwarf KOIs usually have small errors on $R_P/R_*$; instead, errors on planetary radii are usually dominated by errors in the stellar radius \citep[e.g.,][]{Muirhead2012, Muirhead2015}. This has motivated a plethora of follow-up programs and efforts to improve methods to constrain M dwarf radii \citep[e.g.,][]{Rojas-Ayala:2012uq,2014MNRAS.437.2831Z,2014A&A...568A.121N,Newton2015A,Hartman2015}. {\it K2} M dwarfs are statistically closer than \kepler\ M dwarfs, but still have poorly constrained stellar radii \citep[e.g.,][]{2015ApJ...804...10C,2015ApJ...809...25M}. Like stars with known parallaxes \cite[e.g., MEarth;][]{2014ApJ...784..156D}, because we know the (kinematic) distance of \name\ its other parameters are more precisely established (e.g., a 6-7\% error in $R_*$). However, because the 30-minute cadence of  the \kepler{} photometry is comparable to the transit duration ($\simeq$45\,minutes), even {\it K2}'s high precision light curves leave us with an $R_P/R_*$ value that is only well constrained in one direction (1$\sigma$ error is $+26\%$, $-6\%$). Thus, \pname\ represents a case where the limiting factor in planet parameters is the light curve, and not the stellar parameters. 

High-cadence photometry from the ground could significantly improve the planet parameters. While \name\ is quite faint in the optical ($V=15.9$) by moving to the near-infrared ($z=12.8$, $K=10.4$) a modest sized ($\ge2$\,m) telescope could achieve the requisite precision (mmag) with a cadence of $<5$\,minutes. This is sufficient to resolve out the transit duration and likely rule out or confirm the high impact parameter that is present in our transit fit posteriors (Section~\ref{sec:transit}). 

The activity lifetime of an M dwarf is $1.2\pm0.4$\,Gyr for a spectral type M2, rising to $4.5^{+0.5}_{-1.0}$\,Gyr at M4 and $7.0\pm0.5$\,Gyr at M5 \citep{2008AJ....135..785W}. The lack of H$\alpha$ emission seen in the \kepler{} M dwarf planet hosts \citep{Mann2013} of similar spectral type suggests that they are all $> 1\,$Gyr, significantly older than \name. A comparison to other transiting planets gives us some insight into how planets evolve beyond the Hyades' age (650-800\,Myr). 

\begin{figure*}
	\centering
	\includegraphics[width=0.95\textwidth]{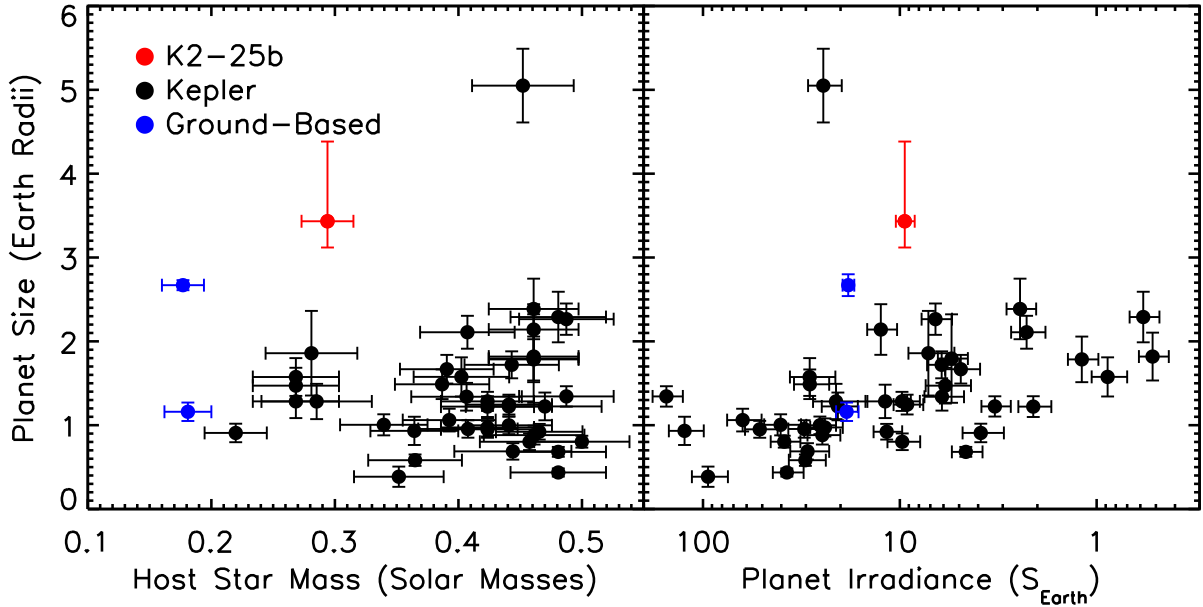} 
	\caption{Planet size as a function of host star mass (left) and planet irradiance (right) for \name\ (red) compared to transiting planets discovered by \kepler{} (black) and from the ground by MEarth (blue). Only planets orbiting stars with host masses $<0.5M_\odot$ and orbital periods $<100$ days are included.  }
	\label{fig:size}
\end{figure*}

In Figure~\ref{fig:size} we show the radius \pname\ as a function of host star mass and planet irradiance compared to transiting planets found by \kepler{} and MEarth with orbital periods $<100$ days and host star masses $M_*<0.5$. Parameters for \kepler{} planets and stars are drawn from \citet{Gaidos2016a}, which are derived in a manner consistent with our own, and parameters for GJ\,1214b and GJ\,1132b are taken from \citet{2013A&A...551A..48A} and \citet{Berta-Thompson2015}, respectively. It is clear that \pname\ is unusually large for its host star mass and orbital period. Only one \kepler{} planet orbiting an $M_*<0.5$ star has a larger radius than \pname\ (KOI 4928.01), but its host star is $\sim$50\% more massive than \name. GJ\,1214b \citep{Charbonneau:2009rt} is the closest match, as the planet is notably smaller, but it also orbits a significantly less massive star.

GJ\,581 has a similar mass ($\simeq0.3M_\odot$) and harbors a nontransiting giant planet, so such large planets are possible. However, it is clear that large planets around very low mass stars are rare. Both \kepler{} and MEarth searched $\simeq1000$ $M_*<0.5M_\odot$ stars \citep{Berta:2013aa,Gaidos2016a} and each found just $\sim$one large planet, while \pname\ was found after searching just $\sim$70 candidate members of the Hyades with $M_*<0.5M_\odot$. Detection biases probably cannot explain the lack of such planets in the \kepler{} sample; close-in, Neptune-sized transiting planets should be obvious in the \kepler{} data, unless they were flagged as eclipsing binaries owing to a short transit duration.

One possible explanation for the large size of \pname\ is that it is evolving under the influence of the environment of its young host star.  M dwarfs, like most stars, pass through a juvenile phase of elevated UV and X-ray emission, flares, and coronal mass ejections.  Models predict this activity to be capable of removing any weakly bound, primordial hydrogen/helium envelopes from rocky planets on close-in orbits \citep{Lammer2014} and are supported by the ejection of a large cloud of neutral hydrogen around the Neptune-mass GJ~436b \citep{Ehrenreich2015}.  \pname\ could represent an early or intermittent phase of planetary evolution where the loss of a distended atmosphere has not yet reached completion.  Detection and characterization of additional planets in young clusters are needed to test such scenarios. While there are only $\sim100$ M dwarfs in the Hyades observed in Campaign 4, many more will be observed in Campaign 13 and even more M dwarfs in Praesepe and Pleiades. Although Praesepe and Pleiades are more distant and hence their M dwarf members are much fainter, the Pleiades is significantly younger \citep[$\sim$110\,Myr.][]{2015ApJ...813..108D}, and planets as large as \pname\ should be quite obvious around even relatively faint M dwarfs. 

The deep transit of \pname\ makes it a useful target for atmospheric characterization. Follow-up of similar-sized objects, such as GJ\,1214b, has mostly suggested hazy, featureless atmospheres \citep[e.g.,][]{2012ApJ...747...35B,2014Natur.505...66K,2014Natur.505...69K}. However, no one has examined the atmosphere of such a young Neptune-sized planet, and it is not known wheather the atmosphere will show features that are no longer present in older counterparts. A high volatile content in the transmission spectrum also might explain \pname's unusual size.

A measurement of \pname's mass would yield constraints on the planet density, possibly providing insight on the unusual size of this planet. Assuming that the planet is Neptune-mass on a near-circular orbit, the Doppler amplitude is expected to be $\sim15$\,\mps. The scatter in our RV measurements is significantly larger than this ($>200$\,\mps) even after accounting for the expected measurement error ($\simeq$50\mps). We adopted a more conservative measurement of the error ($\simeq$150\mps) to account for jitter common to young stars and the untested long-term stability of IGRINS at the 50\mps\ level. 

The RV scatter may also be due to a nontransiting planet. Additional RV measurements could resolve this question. Even if the source of the noise is astrophysical, it might be possible to remove signals that do not follow a Keplerian trend consistent with the orbital period of \pname. The faint optical magnitude and spot-induced jitter are limitations, but \name\ would be an ideal target for monitoring by NIR spectrographs like CARMENES \citep{Quirrenbach2012}, the Infrared Doppler instrument \citep{Kotani2014}, SPIRou \citep{2014SPIE.9147E..15A}, and the Habitable-zone Planet Finder \citep{Mahadevan2012}.

\acknowledgements
A.W.M. was supported through Hubble Fellowship grant 51364 awarded by the Space Telescope Science Institute, which is operated by the Association of Universities for Research in Astronomy, Inc., for NASA, under contract NAS 5-26555. This research was supported by NASA grant NNX11AC33G to E.G.  E.G. was also supported by a International Visitor grant from the Swiss National Science Foundation. We thank Bandit for his useful discussions and encouragement during the writing of this manuscript. 

This work used the Immersion Grating Infrared Spectrograph (IGRINS), which was developed under a collaboration between the University of Texas at Austin and the Korea Astronomy and Space Science Institute (KASI) with the financial support of the US National Science Foundation under grant ASTR1229522, of the University of Texas at Austin, and of the Korean GMT Project of KASI. The IGRINS pipeline package PLP was developed by Dr. Jae-Joon Lee at Korea Astronomy and Space Science Institute and Professor Soojong Pak's team at Kyung Hee University. SNIFS on the UH 2.2-m telescope is part of the Nearby Supernova Factory project, a scientific collaboration among the Centre de Recherche Astronomique de Lyon, Institut de Physique Nucl\'{e}aire de Lyon, Laboratoire de Physique Nucl\'{e}aire et des Hautes Energies, Lawrence Berkeley National Laboratory, Yale University, University of Bonn, Max Planck Institute for Astrophysics, Tsinghua Center for Astrophysics, and the Centre de Physique des Particules de Marseille. Some of the data presented in this paper were obtained from the Mikulski Archive for Space Telescopes (MAST). STScI is operated by the Association of Universities for Research in Astronomy, Inc., under NASA contract NAS 5-26555. Support for MAST for non-HST data is provided by the NASA Office of Space Science via grant NNX09AF08G and by other grants and contracts. This research was made possible through the use of the AAVSO Photometric All-Sky Survey (APASS), funded by the Robert Martin Ayers Sciences Fund. The authors acknowledge the Texas Advanced Computing Center (TACC) at The University of Texas at Austin for providing grid resources that have contributed to the research results reported within this paper. 

The authors wish to recognize and acknowledge the very significant cultural role and reverence that the summit of Mauna Kea has always had within the indigenous Hawaiian community. We are most fortunate to have the opportunity to conduct observations from this mountain.

{\it Facilities:} \facility{IRTF (SpeX)}, \facility{UH:2.2m (SNIFS)}, \facility{Kepler}, \facility{Keck:II (NIRC2)}, \facility{Smith (IGRINS)}


\end{document}